\documentstyle[epsfig,pra,aps,multicol]{revtex}

\begin{document}

\def\sec{\section}
\def\ss{\subsection}
\def\sss{\subsubsection}

\def\del{\partial}

\def\be{\begin{equation}}
\def\ee{\end{equation}}

\def\bea{\begin{eqnarray}}
\def\eea{\end{eqnarray}}

\def\ba{\begin{array}}
\def\ea{\end{array}}

\def\nn{\nonumber}

\def\ben{\begin{enumerate}}
\def\een{\end{enumerate}}

\def\fn{\footnote}

\def\r{\right}
\def\l{\left}

\def\goto{\rightarrow}
\def\comefrom{\leftarrow}
\def\between{\leftrightarrow}

\def\ra{\rangle}
\def\la{\langle}

\def\ack{\acknowledgements}
\def\refs{\references}

\def\bc{\begin{center}}
\def\ec{\end{center}}

\title{Conductance of one-dimensional quantum wires}
\author{K.-I. Imura, K.-V. Pham, P. Lederer and F. Pi\'echon}
\address{Laboratoire de Physique des Solides, 
B\^at. 510, CNRS, Universit\'e Paris-Sud, 91405 Orsay, France}

\date{\today}

\maketitle

\begin{abstract}
We discuss the conductance of quantum wires (QW) in terms
of the Tomonaga-Luttinger liquid (TLL) theory. We use explicitly 
the charge fractionalization scheme which results from the 
chiral symmetry of the model. We suggest that results of the 
standard two-terminal (2T) conductance measurement depend on 
the coupling of TLL with the reservoirs and can be interpreted 
as different boundary conditions at the interfaces. We propose 
a three-terminal (3T) geometry in which the third contact is 
connected weakly to the bulk of TLL subjected to a large bias 
current. We develop a renormalization group (RG) analysis for 
this problem by taking explicitly into account the splitting 
of the injected electronic charge into two chiral irrational 
charges. We study in the presence of {\it bulk} contact
the leading order corrections to the conductance for two 
different boundary conditions, which reproduce in the 
absence of {\it bulk} contact, respectively, the standard 
2T source-drain (SD) conductance $G_{\rm SD}^{(2)}=e^2/h$ 
and $G_{\rm SD}^{(2)}=ge^2/h$, where $g$ is the TLL charge 
interaction parameter. We find that under these two boundary 
conditions for the {\it end} contacts the 3T SD conductance 
$G_{\rm SD}^{(3)}$ shows an UV-relevant deviation from the above
two values, suggesting new fixed points in the ohmic limit.
Non-trivial scaling exponents are predicted as a result of
electron fractionalization.
\end{abstract}
\pacs{71.10.Pm, 72.10.-d, 73.23.-b}

\begin{multicols}{2}

\sec{Introduction}
Interacting electrons in one spatial dimension (1D) 
are one of the best examples of strongly correlated fermionic systems. 
They are  usually discussed in terms of the Tomonaga-Luttinger liquid 
(hereafter TLL). The latter has allowed to discuss in a precise
 fashion the breakdown of the Fermi liquid picture which is a good description 
of interacting electrons in broad band metallic three dimensional systems. 
In 1D, there are no quasi-particles corresponding to a free electron with charge $-e$
and spin $1/2$: the electron Green's function exhibits no quasi-particle pole,
 the density of states at the Fermi level vanishes at the Fermi level, 
and behaves as a power law with non integer exponents as a function of  energy; 
last but not least, spin degrees of freedom are dynamically split from charge degrees 
of freedom. Both propagate at different velocities.

The TLL is usually understood in terms of 
the density fluctuations 
at finite wave vector, and zero wave vector "zero modes". 
Recently, however, taking advantage of the chiral symmetry ,
a new approach of the TLL Hamiltonian  succeeded  
in formulating its physics in terms of 
 generally {\it irrational excitations},
 i.e., excitations which may have dynamically independent irrational charges 
or spin \cite{pham}. This constitutes a generalization 
of the Laughlin {\it fractional} charge excitations which have been observed 
in fractional quantum Hall (FQH)
samples \cite{goldman,noise}. More precisely, 
the irrational excitations  have been shown  to be eigenstates 
of the TLL hamiltonian in the chiral representation.
Their wave functions  are formally isomorphic to Laughlin 
wave functions for FQH states.
We consider hereafter a spinless TLL for simplicity.
The irrational charges carried by the irrational excitations are
created in chiral pairs with one excitation moving to the left and
the other to the right; the charges carried by each possible pair
of states form a 2D manyfold $(Q_{+},Q_{-})$, where
$Q_{\pm}={1\over 2}(N\pm gJ)$
with $N$ and $J$ being integers having the same parity:
$(-1)^{N}=(-1)^{J}$. 
$N$ and $J$ are standard zero-mode quantum numbers associated,
respectively, with the total charge and the persistent current of the 
system.
\cite{hal}
$g$ is the so-called TLL parameter, which contains all the relevant
information of the electron-electron interaction.
$g$ takes the value, $0<g<1$ for a standard repulsive interaction,
whereas $g=1$ for non-interacting 1D fermions.

The main question to be asked is 
whether or not these irrational charge
excitations are observable. Consider an electron incident in the 
middle of TLL either of infinite length or sufficiently far away from
the boundaries so that the chiral symmetry be preserved.
The injected electron splits into two eigen excitations
which have irrational charges $Q_+=(1+g)/2$, $Q_-=(1-g)/2$
(or vice versa),
and propagate in opposite directions.
This picture is quite reminiscent of a three-terminal 
conductance measurement
in which the third terminal
is attached to the middle of QW 
connected to the source (S) and drain (D) (See Fig. 1).
The main findings of this paper are that the standard conductance
measurement done in this geometry does provide some information
on the charge fractionalization.
More precisely, we consider a source-drain (SD) conductance
$G_{\rm SD}^{(3)}$,
under a large bias current $I_{\rm bias}$ circulating through
the QW between S and D,
and
in the {\it presence} of bulk-injected current $i_{\rm bulk}$.
The third terminal, or a bulk contact, at voltage $V$
is then connected by an ohmic wire either to the S or to the D.
We consider the case of small $i_{\rm bulk}$, i.e.,
the case in which $-eV$ is close to the chemical potential of
the reservoir from which $i_{\rm bulk}$ is provided.
We find under these circumstances
$G_{\rm SD}^{(3)}$
is subjected to a change 
characterized by unusual scaling exponents which take different 
values depending on
$G_{\rm SD}^{(2)}$ 
in the {\it absence} of $i_{\rm bulk}$.
In Sec. III we will explain in length that different values of
$G_{\rm SD}^{(2)}$ can be interpreted as different
boundary conditions at the {\it end} contacts to S and D. 
In this paper, we highlight two specific boundary conditions
which correspond to 
$G_{\rm SD}^{(2)}=e^2/h$ ($\leftrightarrow$ boundary condition A)
and
$G_{\rm SD}^{(2)}=ge^2/h$ ($\leftrightarrow$ boundary condition B).
In these terms we found 
a non-trivial scaling exponent $(2\Delta)^2-1$
(See Eq. (\ref{result1}))
under a specific boundary condition A,
whereas a standard scaling exponent $2\Delta-1$
(See Eq. (\ref{result2}))
for boundary condition B,
which is simply related to the anomalous scaling
dimension  $\Delta$ of the TLL electron operator.
This result
is a remarkable consequence of electron fractionalization
under a stationary bias current $I_{\rm bias}$.

A motivation of this paper is therefore closely related to the
so-called "conductance puzzle" of the QW:
an apparent contradiction among different theoretical and
experimental results for the two-terminal (2T) conductance 
$G_{\rm SD}^{(2)}$
in the ohmic regime. \cite{safi1,maslov,pono1}
In spite of the theoretical prediction that
the interaction should renormalize the conductance as
$G_{\rm SD}^{(2)}=ge^2/h$, \cite{apel} one of the first conductance 
measurements on a QW by Tarucha et al. \cite{tarucha}
has found a non-renormalized universal conductance
$G_{\rm SD}^{(2)}=e^2/h$ for an interacting system.
On the contrary, in the case of the fractional quantum Hall 
(FQH) edge mode, an example of a chiral TLL \cite{wen},
the Hall conductance, usually measured in a four-terminal geometry,
is  maximally renormalized at a topological number:
$G_{\rm H}=\nu e^2/h$ \cite{tsui}.

It turns out that a rapidly growing number of experimental results
is now available, on the conductance  
of QW \cite{yacoby,picciotto1,picciotto2} 
and of carbon nanotubes \cite{bockrath,laughlin,kasumov}. 
Carbon nanotubes \cite{saito,dekker} 
have been expected since their discovery
\cite{iijima}
to be  ideal 1D quantum wires.
The single-walled nanotubes (SWNT) have four conducting
channels indicating an expected quantized conductance:
$G_{\rm SD}=4e^2/h$ \cite{egger1}. The ballistic transport in carbon nanotubes 
was first observed in only one channel of
multi-walled nanotubes (MWNT) with
$G_{\rm SD}\sim 2e^2/h$
\cite{deheer}.
The temperature and bias voltage dependence of
the conductance reported in Ref. 
\cite{bockrath} do suggest that this system is a strongly correlated
1D electronic liquid.  The TLL theory for carbon nanotubes \cite{gogolin} 
has suggested an interaction parameter $g$ in the range $0.2-0.3$.

Recent data on the conductance of QW and carbon nanotubes display 
a variety of results. In Ref. \cite{yacoby,picciotto1}, significant
deviations from the quantized value $e^2/h$ were observed. In Ref.
\cite{laughlin},
the observed conductance exhibits fluctuations versus Fermi energy approaching
the unrenormalized theoretical value $4e^2/h$ as the temperature is lowered.
On the other hand, in the experiment by Kasumov et al.
\cite{kasumov} the
isolated SWNT exhibits a resistance which saturates at low temperature 
(in the presence of a sufficiently intense magnetic field)
to a factor of about 0.25 times the expected unrenormalized value. 
In our route to suggesting experimental ways of observing irrational excitations,
we had to spend some time trying to understand this variety of results.
Our understanding, as explained in the body of this paper, 
is that two terminal conductance  measurements should indeed 
display this variety of results, which may be understood as 
expressing a variety of boundary conditions at the {\it end} contacts.

The main effort of this paper is devoted to studying more involved 
experimental geometries than two terminal ones. 
To our knowledge the only way to describe injection
of electrons in a QW through a weak {\it bulk} contact is to resort
to the irrational excitation picture,
which should be taken into account in
the theoretical description of this process.
Thus, examining electron injection, from one (various) weak bulk 
contact(s), in a QW connected to reservoirs at its ends,  
should lead to specific experimental predictions, 
as we argue in the body of this work. 
We were stimulated in that direction by the 
work of Chamon and Fradkin \cite{chamon}. 
That work deals with the FQH effect and examines the conductance of a 
FQH bar (See Sec. III-C).
In the case of the non-chiral liquid, one cannot 
easily manufacture electrical contacts which inject electrons only 
in one chiral eigen-mode of the TLL Hamiltonian density, 
so that one cannot use the results of \cite{chamon}.
One must actually solve the problem of the 
non-chiral TLL with many leads.

We derive new scaling exponents 
associated with 
the currents injected from {\it bulk} contacts. 
More generally we discuss effects 
which are derived using the irrational excitation picture,
allowing measurements of $g$ through multi-terminal conductance
measurements. 
Deviations from the unrenormalized perfect conductance value are 
predicted
in the ohmic limit.
In the renormalization group (RG) picture
this naturally suggests a possibility of 
new intermediate fixed points.
We have not been able, though, to prove that our results are 
{\it unique}
predictions of the irrational excitation scheme, 
so that experimental observation of, say, the new scaling exponent 
mentionned above would be at best a plausibility
argument in favour of this scheme. 
The discussion of shot-noise experiments is also
left for a future publication.
\cite{future}

This paper is organized as follows: 
Sec. II describes the model we are studying. That section
is an attempt to clarify the notion of the chemical potential 
for (interacting) eigen-modes, 
as opposed to the chemical potential for bare electrons.
Sec. III discusses equilibration of 1D conductors 
with the 3D reservoirs: in  actual experiments, 
which are the particles which equilibrate with the reservoirs:
the bare electrons, or the eigen-modes of the TLL ? 
This analysis allows a physical interpretation of
our boundary conditions.
In Sec. IV we solve our {\it chiral} electric circuit equations
using the technology developed in Sec. III.
Sec. V is devoted to the RG analysis of a {\it bulk} contact.
Using our {\it fractional} scaling analysis,
we demonstrate that a non-trivial scaling exponent appears
as a result of a specific boundary condition (boundary condition A). 
Sec. VI is a generalization to many bulk
contacts. Some remarks on the application to SWNT and MWNT
will be found in Sec. VII. 
Our conclusions are discussed in Sec. VIII.

\section{Model, notations and chiral densities}
Of importance to us in this paper is the distinction to be made
between bare chiral electron densities and 
eigen-modes chiral densities of the TLL.
The bare chiral electron densities $\rho_{\pm}^{\rm (0)}(x,t)$
correspond to the densities of
electrons created either at the left or right Fermi points of a 
\textit{non-interacting} system.
The total electronic density $\rho(x,t)$
and the current density $j(x,t)$ are related to
$\rho_{\pm}^{\rm (0)}(x,t)$ as
$\rho(x,t)=\rho_{+}^{\rm (0)}+\rho_{-}^{\rm (0)}$,
$j(x,t)=v_{F}\left(\rho_{+}^{\rm (0)}-\rho_{-}^{\rm (0)}\right)$.
In the \textit{non-interacting} system
the bare chiral electron densities are indeed two independent
eigenmodes of the system:
$\rho_{\pm }^{\rm (0)}(x,t)=\rho_{\pm }^{\rm (0)}(x\mp v_{F}t)$.
It is however no longer true
in the TLL. In the interacting system
the left and right-moving electrons
of the non-interacting system are strongly coupled together;
accordingly, the bare electronic densities are no longer
chiral.
In order to clarify this point
we consider the harmonic hamiltonian density
of the spinless TLL,
\be
{\cal H}_{\rm TLL}={u\over 2}
\left[{1\over g}
\l(\frac{\partial\Phi(x,t)}{\partial x}\r)^2
+g\ \Pi(x,t)^2\right],
\label{ham}
\ee
where we have introduced the standard phase field
$\Phi$ related to the electron density by:
$\rho(x,t)={1\over\sqrt{\pi}}
{\partial\Phi(x,t)\over\del x}$,
and its conjugate canonical momentum $\Pi(x,t)$.
$u=v_F/g$ is the dressed velocity.
Note also that the continuity equation shows
that the current density is simply:
$j(x,t)=-{1\over \sqrt{\pi}}{\partial\Phi(x,t)\over\del t}$.
The stationary components of $\rho(x,t)$ and
$j(x,t)$ are the zero modes:
$N=\int_{-L/2}^{L/2}\rho(x,t)$, $J=\int_{-L/2}^{L/2}j(x,t)$,
which obey the fermionic
selection rule $(-1)^N=(-1)^J$
\cite{hal}.
Using Hamilton equations:
$ug\ \Pi(x,t)={\partial\Phi(x,t)\over\del t}$,
$\frac{u}{g}\ {\partial^2\Phi(x,t)\over\del x^2}
={\partial\Pi (x,t)\over\del t}$,
one finds immediately
\be
\l({\del\over\del x}\mp\frac{1}{u}{\del\over\del t}\r)
\l[{\partial\Phi (x,t)\over\del x}\pm g\ \Pi (x,t)\r]=0.
\ee
This shows that
$\rho_{\pm }(x,t)=\frac{1}{2\sqrt{\pi}}
\l[{\partial\Phi(x,t)\over\del x}\pm g\Pi(x,t)\r]$
are indeed chiral eigenmodes of the system:
$\rho_{\pm}(x,t)=\rho_{\pm}(x\mp ut)$.
Observing that
$\rho(x,t)=\rho_{+}+\rho_{-}$,
$j(x,t)=u\l(\rho_{+}-\rho_{-}\r)$,
one concludes that
$\rho_{+}$ and $\rho_{-}$ correspond to a 
different decomposition of the total density into 
chiral densities from the non-interacting case.
These eigenmode chiral
densities mix both left and right moving electrons,
since the bare chiral densities
(obtained when $g=1$) are:
$\rho^{\rm (0)}_{\pm }=\frac{1}{2\sqrt{\pi }}
\l[{\partial\Phi(x,t)\over\del x}\pm\Pi(x,t)\r]$.
In terms of these eigenmode chiral densities obeying 
the anomalous
Kac-Moody commutation relations, the hamiltonian 
density splits into
two commuting chiral parts:
${\cal H}=\frac{\pi u}{g}\rho ^{2}_{+}+
\frac{\pi u}{g}\rho _{-}^{2}
={\cal H}_{+}+{\cal H}_{-}$.
The stationary component of $\rho_{\pm}(x,t)$ are
nothing but the chiral charges
$Q_\pm=\int_{-L/2}^{L/2}\rho_\pm(x,t)$
\cite{fn1}.
It is convenient to introduce a vector notation for the chiral 
densities.
The dressed eigenmode density
$\vec{\rho}=\l[\ba{r}\rho_+\\ \rho_-\ea\r]$
is related to the bare density
$\vec{\rho}^{\rm (0)}=\l[\ba{r}\rho_+^{\rm (0)}\\
\rho_-^{\rm (0)}\ea\r]$
in the matrix equation as
$\vec{\rho}={\bf \Omega}\ \vec{\rho}\ ^{\rm (0)}$,
where the matrix 
${\bf \Omega}={1\over 2}\l[\ba{rr}
1+g & 1-g \\
1-g & 1+g
\ea\r]$ characterizes the fractionalization of electronic
charge $-e$.

In the absence of applied external voltage
$V_S-V_D$ the average current $I=\la j(x,t) \ra$
is zero. In order to drive a net current
through the sample,
let us allow for independent variations
of the left and right bare chemical potentials.
The possibility to adjust them independently
expresses the chiral separation of TLL.
This is accomplished
by adding a chemical potential to the hamiltonian.
But once again a distinction should be made 
between {\it bare} chemical potentials
$\vec{\mu}^{\rm (0)}=
\l[\ba{r}\mu_+^{\rm (0)}\\ \mu_-^{\rm (0)}\ea\r]$
corresponding
to a variation of the bare electron densities
and {\it eigenmode} chemical potentials
$\vec{\mu}=\l[\ba{r}\mu_+\\ \mu_-\ea\r]$
corresponding to the eigen-mode
chiral densities.
More precisely,
$\vec{\mu}^{\rm (0)}$ and $\vec{\mu}$
are defined, respectively, by minimizing
\bea
{\cal H}_{\rm TLL}-\mu^{(0)}_{+}\rho^{(0)}_{+}-
\mu^{(0)}_{-}\rho^{(0)}_{-}
&=&{u\pi\over 2g}\vec{\rho}^{\ (0)T}
{\bf \Omega}^2\vec{\rho}\ ^{\ (0)}-
\vec{\mu}^{\ (0)T}\vec{\rho}^{\ (0)},
\nn \\
{\cal H}_{\rm TLL}-\mu_{+}\rho_{+}-\mu_{-}\rho_{-}
&=&{u\pi\over 2g}\vec{\rho}\ ^T\vec{\rho}-
\vec{\mu}^{T}\vec{\rho}.
\nn 
\eea
Completing the square densities, 
one finds
$\vec{\mu}^{\ (0)}=
{u\pi\over g}{\bf \Omega}^2\la\vec{\rho}^{\rm (0)}\ra$,
$\vec{\mu}={u\pi\over g}\la\vec{\rho}\ra
={u\pi\over g}{\bf \Omega}\la\vec{\rho}^{\rm (0)}\ra$
\cite{safi1,alek1}.
Comparing the two expressions,
the relation between bare and dressed chemical potential
is found to be 
$\vec{\mu}^{\rm (0)}={\bf \Omega}\vec{\mu}$.
Note that $\vec{\mu}^{\rm (0)}$ and $\vec{\mu}$
obey the same linear transformation as the one for
$\vec{\rho}^{\rm (0)}$ and $\vec{\rho}$ except that
the roles of bare elctrons and of the chiral eigenmodes 
are exchanged.

It would be worth mentioning here that
in the four-terminal conductance measurement
by Picciotto et. al.
the resistance data (Fig. 3 of \cite{picciotto2}) shows that
the chemical potential $\mu$
coupled to additional probes (probe A and B) 
is neither bare nor dressed chiral chemical
potentials.
Instead the additional probes seem to be
coupled almost equally to both chiralities;
$\mu$ is coupled to the total density:
we should rather minimize the hamiltonian density
${\cal H}_{\rm TLL}-\mu\rho$
to find $\mu={\mu_+ +\mu_-\over 2}$.
If the electronic transport through the
conductor is perfectly balistic,
this chemical potential $\mu$ is uniform throughout
the conductor, which explains the data of
Picciotto et. al.

In the presence of an electric field the chemical 
potential becomes
an electrochemical potential and one may 
introduce the following chiral voltages: 
$\vec{\mu}^{(0)}=
-e\vec{V}^{(0)}=\l[\ba{r}-eV_+^{(0)}\\-eV_-^{(0)}\ea\r]$,
$\vec{\mu}=
-e\vec{V}=\l[\ba{r}-eV_+\\-eV_-\ea\r]$.
They are related, of course, via the
relation $\vec{V}={\bf \Omega}\vec{V}^{\rm (0)}$.
Thus the total current
$I=-e\la j(x,t)\ra$ can be expressed either
in terms of the bare voltages
$I=-eu\ [1, -1]\vec{\rho}=
{e^2\over h}\ [1, -1]\ \vec{V}^{\ (0)}$
or in terms of the dressed eigenmode voltages as
$I=-ev_F\ [1, -1]\vec{\rho}^{\rm (0)}=
g\ {e^2\over h}\ [1, -1]\ \vec{V}$,
where we have used $[1, -1]\ {\bf \Omega}=g\ [1, -1]$.
Note that we are working in the unit
where $\hbar=1$.
These relations together with
$\vec{V}={\bf \Omega}\vec{V}^{\rm (0)}$
play a central role in later
sections when the sample is connected to the
reservoirs through various boundary conditions.

We have summarized the bosonized formulation of
TLL as well as its response to external
electric field by emphasizing the difference between bare and 
eigen-mode chiral densities. 
Let us now turn to a discussion of its implications on the 
transport through TLL.

\section{Boundary conditions at the end contacts
--- screening and equilibration}

In the approach developped by Landauer and B\"uttiker
for non-interacting electrons,
the chiral chemical potentials
$-eV_+$, $-eV_-$ of the bulk sample are equilibrated
with that of the reservoir from which the electrons are
injected \cite{landauer}:
$V_S=V_+$, $V_D=V_-$.
Whereas in the bulk the total current $I$ is
related to $V_+$, $V_-$ 
as $I={e^2\over h}(V_+ -V_-)$.
Thus the above boundary condition ensures 
the 2T SD conductance, usually defined as 
$G_{\rm SD}^{(2)}=I/(V_S-V_D)$, to be
given by the standard unit conductance:
$G_{\rm SD}^{(2)}=e^2/h$.

Let us now switch on the interaction.
As we have seen in the last section,
the conductance of the system as measured against either the bare
or eigen-mode voltage yields therefore different values,
e.g.,
if the conductance is measured 
against $V_+-V_-$, this gives the conductance $ge^2/h$
in the bulk,
which is reminiscent of the four-terminal measurement in
FQH bar
\cite{tsui}.
In the  case of the non-chiral liquid, 
a four-terminal measurement analogous to
the one in Ref. \cite{tsui} is difficult to realize,
since one cannot easily manufacture 
electrical contacts which are coupled 
only to one of the chiral eigenmodes of TLL Hamiltonian.
In the experiment by Picciotto et. al. 
the resistance data (Fig. 3 of \cite{picciotto2}) shows that
the voltage probes (probe A and B) are coupled almost
equally to both chiralities.
On the other hand,
if the conductance is measured, under certain circumstances,
against $V^{\rm (0)}_{+}-V^{\rm (0)}_{-}$,
then it gives $e^2/h$.
We believe that
the value taken by the conductance,
when the bulk sample is connected to the current reservoirs,
is, a matter of coupling between the sample and the reservoirs.
In the followoing
we will formulate this in a more systematic way, i.e.,
in the form of boundary conditions at the {\it end} contacts.

\ss{Equilibration with bare electrons --- Screening by a metallic gate}
Let us first consider the boundary condition
discussed in refs. \cite{safi1,maslov,pono1,safi2,alek2,egger2}.
This boundary condition has been labelled a
``radiative'' boundary condition in Ref. \cite{egger2}.
With this boundary condition
the particles emitted by the left reservoir
are then in equilibrium with the {\it bare} electrons:
\be
\l\{\ba{l}
V_S=V_+^{\rm (0)}=[1\ 0]\ {\bf \Omega}\ \vec{V}\\
V_D=V_-^{\rm (0)}=[0\ 1]\ {\bf \Omega}\ \vec{V}
\ea\r.,
\label{bc0}
\ee
where the matrix ${\bf \Omega}$ has been defined as
\be
{\bf \Omega}={1\over 2}\l[\ba{rr}
1+g & 1-g \\
1-g & 1+g
\ea\r].
\label{omega}
\ee
Let us first recall that 
the conductance defined with this boundary condition
is indeed $e^2/h$ independently of $g$.
Recall the relation between bare and dressed
voltages: $\vec{V}={\bf \Omega}\vec{V}^{\rm (0)}$.
The bias voltage
$V_S-V_D=[1, -1]\ \vec{V}\ ^{\rm (0)}$
can be written as
$V_S-V_D=[1, -1]\ {\bf \Omega}\ \vec{V}
=g\ [1, -1]\ \vec{V}$,
where we have used $[1, -1]\ {\bf \Omega}=g\ [1, -1]$.
The total current $I$ can be expressed either
in terms of the bare bias voltages
$I={e^2\over h}\ [1, -1]\ \vec{V}^{\ (0)}$
or in terms of the dressed eigenmode voltages as
$I=g\ {e^2\over h}\ [1, -1]\ \vec{V}$.
It then follows that the conductance is
given by $G_{\rm SD}^{(2)}=e^2/h$ independently of $g$.

Another important remark is that
this boundary condition requires the existence
of a metallic gate along the 1D sample
\cite{safi2,egger2}.
When a certain amount of charge $Q$ is injected
from the reservoir through an {\it end} contact,
the TLL system cannot screen this charge completetely
because of this boundary condition, instead
$Q_{\rm TLL}=-(1-g^2)Q$ is induced in TLL.
In order for the charge conservation
$Q+Q_{\rm TLL}+Q_{\rm gate}=0$ to be satisfied
there needs to exist a metallic gate
providing for a screening charge
$Q_{\rm gate}=-g^2Q$ 
\cite{safi2,egger2}.
The existence of metallic gate also explains
a short-range interaction in 1D quantum wire
which ensures a finite parameter $g$ of TLL model.
In the experiment by Tarucha et. al. \cite{tarucha}
the 1D sample was indeed screened by the metallic
gate.

\ss{Equilibration with dressed eigenmodes --- No screening gate}
Knowing that the boundary condition (\ref{bc0})
requires the existence of metallic gate
one naturally asks the question
what will be the corresponding boundary conditions
in the absence of screening by a metallic gate.
TLL with finite $g$ (long-range interaction
cut off by the finite length of the sample and
the width of the tube) without screening gate
is indeed realized in a SWNT \cite{egger1}.
The boundary condition which ensures
$Q+Q_{\rm TLL}=0$ with no reference to the metallic gate is,
in fact, $V_{+}=V_{S}$, $V_{-}=V_{D}$.
The dressed eigenmode voltages are equilibrated with
those of the reservoirs, i.e.,
the particles emitted from the left (right)
reservoir are moving to the right (left) and are in 
equilibrium with the
right (left)-moving eigenstates of the TLL.
If this naive picture is indeed the case, the conductance 
in terms of the voltage difference between the reservoirs, 
is obtained immediately from the relation
$I=g\ {e^2\over h}\ [1, -1]\ \vec{V}$,
i.e., $G_{\rm SD}^{(2)}=ge^2/h$.
The conductance is fully renormalized
\cite{future}.

\ss{Intermediate possibilities}
Are there intermediate possibilities, i.e.,
situations where
neither bare electrons nor eigenstates
are in equilibrium with the reservoir?
In the case of FQH edge mode, i.e.,
if one replaces in the above discussion the 1D TLL sample by a FQH bar
at bulk filling factor $\nu$, 
it was shown in Ref. \cite{chamon}
that depending on the number $N_L$ ($N_R$) of
left (right) strong point like contacts 
with the left (right) reservoir 
the {\it SD conductance of a FQH bar}
$G_{\rm SD}$ varies from $G_{\rm SD}=e^2/h$
to $G_{\rm SD}=\nu e^2/h$.
The first case (boundary condition A) in which the 2T conductance
is not fractional but integral ($G_{\rm SD}^{(2)}=e^2/h$) 
corresponds to a single tunneling point contact:
$N_L=N_R=1$.
The second case (boundary condition B) where the {\it end} contacts are in equilibrium
with dressed eigenmodes corresponds to an infinite
number of tunneling point contacts between each
reservoir and the sample: $(N_L,N_R)\goto (\infty,\infty)$.
The third possibility (boundary condition C) arises in between,
when there is a finite number of tunneling quantum point 
contacts.

Turning to the case of the quantum wire,
which of these three cases is relevant
experimentally?
A first model is that of the inhomogeneous TLL, in which 
the reservoirs are modelized as TLL at $g=1$ (1D free fermions). 
\cite{safi1,maslov,pono1}
We observe that this model implicitly describes
a transmission of the current through a single point contact. 
Furthermore it implies the presence of a screening gate.
This confirms that a single contact between the reservoirs 
is not enough for equilibration of the 1D sample 
with the reservoirs as we surmised, yielding
therefore an unrenormalized conductance. 
A second series of explanations rely on the assumption 
that the conductance
measured is the ratio \( I/V_{\rm loc} \) of the current
to a so-called
total local field, which sums a contribution from the 
external electrical
field plus some response of the TLL, rather than 
the ratio of the current
to the external potential as usual
\cite{kawab}.
It is then shown by a diagrammatical 
analysis that the local field is 
\( E_{\rm loc}=E_{\rm ext}/g \)
\cite{finkel}.
The ratio \( I/V_{\rm ext} \) being assumed as in the initial work by
Kane and Fisher \cite{kane} to be equal to \( g \), 
this yields
an unrenormalized conductance \( I/V_{\rm loc}=1 \). That analysis is
doubtful because experimentally the external field is fixed which
means that  one really measures  the ratio \( I/V_{\rm ext} \).
By contrast, in our approach what is measured is indeed the ratio
\( I/V_{\rm ext}=1 \), although we agree on the relation 
\( E_{\rm loc}=E_{\rm ext}/g \).
Note that this follows from  elementary
considerations on chiral modes (see section II).

We stress that 
although the boundary condition A seems to be realized
in the experiments when a screening gate is present,
this does not bar the possibility that, in different 
set-ups, boundary condition B or C might apply. 
In the experiment by Kasumov et. al.
\cite{kasumov}
the isolated SWNT sample $ST_1$ exhibits a resistance $R$
which saturates at high temperatures to $R\sim 25 {\rm k\Omega}$
(see Fig. 2B of Ref. \cite{kasumov}).
This corresponds to the conductance
$G_{\rm SD}^{(2)}\sim e^2/h$,
which is smaller by a factor $1/4$
than $G_{\rm SD}^{(2)}=4e^2/h$ expected from the boundary condition
(\ref{bc0}). 
Note that this experiment was done in the absence of
screening gate, where the boundary condition B might apply.
Since $g$ is typically in the range $0.2$ to $0.3$,
the observed conductance
$G_{\rm SD}^{(2)}\sim e^2/h$ is indeed reminiscent of the renormarized
conductance: $G_{\rm SD}^{(2)}=4ge^2/h$.

The inhomogeneous TLL model \cite{safi1,maslov,pono1}
assumes by construction an injection of current 
through a single end
contact for each reservoir. 
But the current might be injected through
{\it bulk} contacts in addition to being injected through {\it end} contacts.
In further sections of this paper we address the issue of
{\it bulk} contacts using the perpurbative RG analysis for bulk tunneling.

\section{One bulk contact in the presence of a large bias current}
In the following we will consider a model
where the sample is connected to the reservoir
not only through the {\it end} contacts but also
through some {\it bulk} contacts.
For the {\it end} contact,
depending on the presence or absence of screening gate,
the boundary conditions (A) or (B) will be applied.
We will mainly focus on the case where the two {\it end} contacts
impose a large stationary current $I_{\rm bias}$ in the quantum wire.
In this case the additional current injected through the {\it bulk} contact
can be treated as a small perturbation.
We will develop later a RG analysis which is compatible
with fractionalization scheme, which will tell us the scaling 
behavior of this bulk injected current.
In order to utilize the results of this fractional RG
analysis, we first have to derive and solve the equations
which describe our {\it chiral} electric circuits.
We will see that the existence of a large stationary current
as well as different boundary conditions
play an important role in finding a new exponent
in the conductance formula.

A motivation of our work is based on the findings of Ref.
\cite{pham}. Following the latter work, an electron 
injected from a normal lead 
into a TLL splits into two eigen excitations which are {\it fractional}
(in fact, for most repulsive interaction strengths, the correct 
term would be {\it irrational}).
Namely, the two chiral eigen excitations 
have charges $(1+g)/2$, $(1-g)/2$, and they propagate in opposite 
directions with the corresponding chiral mode velocities.

Various authors have discussed the observability of those irrational
excitations $ge$ \cite{pham,kane,glaz,trau,bena,pono2}.
The obvious suggestion is based on electric current noise 
measurements \cite{pham}.
However the results of Ref. \cite{chamon} suggest that
the parameter $g$ might also be measured in a conductance 
experiment with many contact points, and/or possibly 
different bias voltages. This is what we proceed
to do below, in various situations.

Let us first consider a single {\it bulk} contact in the middle of
the sample. The {\it bulk} contact has the following two effects
(see Fig. 1).
\ben\item
Tunneling of Laughlin quasiparticles \cite{pham}
between the two chiral modes. This is due to the \rm backward
scattering between $+k_{\rm F}$ and $-k_{\rm F}$ electrons.
The total charge is conserved in this process 
($N=0, J=\pm 2$).
When a macroscopic number of electrons are
involved in this process,
a current $i_{\rm back}$ is \rm back-scattered 
from $(+)$-chirality to $(-)$-chirality.

\item
Tunneling of electrons into or out of
the bulk sample through the {\it bulk} contact,
i.e., injection or ejection of electrons
throught the {\it bulk} contact.
The total charge is increased or decreased by 1:
$N=\pm 1,J=\pm 1$, 
respectively, for the tunneling of 
$+k_{\rm F}$ ($-k_{\rm F}$) electrons.
When a current $i_{\rm bulk}$ is injected through the {\it bulk} contact 
into the sample,
it splits into two parts $i_+$ and $i_-$
corresponding, respectively, to
an eigen modes with $+(-)$-chirality: $i_{\rm bulk}=i_+ +i_-$.
\een

\vspace{-2.0cm}
\begin{figure}[h]
\epsfig{figure=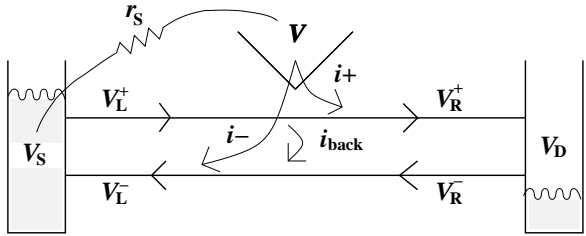,width=6.5cm,angle=-90}
\vspace{-1.0cm}
\caption{A {\it bulk} contact in the presence of
a large stationary current $I_{\rm bias}$.
The chiral symmetry of the TLL is {\it not} destroyed by the
{\it bulk} contact.
---
At the {\it bulk} contact a current $i_{\rm back}$ is \rm back-scattered
from $(+)$-chirality to $(-)$-chirality due to the
tunneling of Laughlin quasiparticle.
A net current $i_{\rm bulk}$ is injected through the {\it bulk} contact
into the sample.
As soon as it is injected into the TLL sample, 
it splits into two parts $i_+$ and $i_-$
corresponding, respectively, to
one of the TLL eigenmodes with $+(-)$-chirality.}
\end{figure}

In order to see how the {\it bulk} contacts influence the two-terminal 
conductance of the system, we develop below
the voltage drop equations, i.e., a set of equations
which determine the chemical potential of the
system on each side of the {\it bulk} contacts.
In the presence of
1. \rm back-scattering,
2. electron injection (ejection),
the voltage drop equation reads
\be
\l\{\ba{l}
i_{+}-i_{\rm back}=g{e^2\over h}(V_{R}^{+}-V_{L}^+)\\
i_{-}+i_{\rm back}=g{e^2\over h}(V_{L}^{-}-V_{R}^-)
\ea\r.,
\label{drop1}
\ee
where $V_L^{\pm}$ and $V_R^{\pm}$ are
the electrostatic potentials of the
eigen modes on each side of the {\it bulk} contact.
Using the vector notation
$\vec{i}_{\rm bulk}=\l[\ba{r}
i_+\\
i_-
\ea\r]$,
$\vec{V}_{L}=\l[\ba{r}
V_L^{+}\\
V_L^{-}
\ea\r]$,
$\vec{V}_{R}=\l[\ba{r}
V_R^{+}\\
V_R^{-}
\ea\r]$,
the voltage drop equations can be rewritten
in a simpler form and can be treated in a systematic way,
\be
\vec{V}_R-\vec{V}_L
={{\bf \sigma_z}\vec{i}_{\rm bulk}\over ge^2/h}
-{i_{\rm back}\over ge^2/h}\l[\ba{r}1\\1\ea\r],
\label{drop2}
\ee
where
${\bf \sigma_z}=\l[\ba{rr}
1 & 0\\
0 & -1
\ea\r]$.
For later convenience we introduce
the following decomposition of $\vec{i}_{\rm bulk}$:
\be
\vec{i}_{\rm bulk}=i_{+k_{\rm F}}{\bf \Omega}\l[\ba{r}1\\ 0\ea\r]
+i_{-k_{\rm F}}{\bf \Omega}\l[\ba{r}0\\ 1\ea\r],
\label{dec}
\ee
where the matrix ${\bf \Omega}$ was defined in Eq. (\ref{omega}).
$i_{+k_{\rm F}}$ and $i_{-k_{\rm F}}$ are, respectively, 
the currents injected from the
{\it bulk} contact through
$\Psi_{+k_{\rm F}}\sim
e^{i\sqrt{\pi}\ [1\ 0]\ {\bf \Omega}
\l[\ba{r}\theta_+\\ \theta_-\ea\r]}$,
or through
$\Psi_{-k_{\rm F}}\sim
e^{i\sqrt{\pi}\ [0\ 1]\ {\bf \Omega}
\l[\ba{r}\theta_+\\ \theta_-\ea\r]}$,
where
$\theta_\pm=\theta\mp{\phi\over g}$ defined in terms of
$\phi=\Phi(x=0),\theta=\Theta(x=0)$
are the values of chiral fields
at the {\it bulk} contact ($x=0$).
Noticing that $[1, 1]\ {\bf \Omega}=[1, 1]$,
one can easily check that
the total bulk injected (ejected) current is indeed
$i_{\rm bulk}=[1, 1]\vec{i}_{\rm bulk}=i_{+k_{\rm F}}+i_{-k_{\rm F}}$.

The total current $I_L$, $I_R$
on each side of the {\it bulk} contact
is related to $\vec{V}_{L}$, $\vec{V}_{R}$
via the equations:
$I_L={e^2\over h}[1, -1]\ {\bf \Omega}\vec{V}_L
=g{e^2\over h}[1, -1]\vec{V}_L$,
$I_R={e^2\over h}[1, -1]\ {\bf \Omega}\vec{V}_R
=g{e^2\over h}[1, -1]\vec{V}_R$.
Let us first focus on
\be
I_R-I_L=g{e^2\over h}[1, -1]\l(
{{\bf \sigma_z}\vec{i}_{\rm bulk}\over ge^2/h}
-{i_{\rm back}\over ge^2/h}\l[\ba{r}1\\1\ea\r]\r)=i_{\rm bulk},
\ee
where we have used (\ref{drop2}).
The physical meaning of this
simplest equation is far from uninteresting.
\ben\item
The total current is conserved
on each side of the {\it bulk} contact,
i.e., $i_{\rm back}$ does not appear in $I_L-I_R$, 
reflecting the fact that the \rm back-scattering 
conserves the total charge.
\item
Even though an electron injected from the
{\it bulk} contact (normal lead) splits into two fractional excitations
propagating in the opposite directions, leading to 
$i_+ -i_-=[1, -1]\vec{i}_{\rm bulk}=
g\ (i_{+k_{\rm F}}-i_{-k_{\rm F}})$
(see Eq. (\ref{dec})),
the total current added by $\vec{i}_{\rm bulk}$ is 
$i_{\rm bulk}$ independently of $g$.
\een
Now we connect both ends of the sample to the 
reservoirs through either of the two boundary
conditions discussed in section III.

\ss{One bulk contact with end contacts
in equilibrium with bare electrons}
Let us first consider the case where the
sample is connected to the reservoirs
via the boundary condition (\ref{bc0}).
Using the voltage drop equation (\ref{drop2}),
the boundary condition (\ref{bc0})
at the {\it end} contacts can be written as
\bea
V_S&=&[1\ 0]\ {\bf \Omega}\ \vec{V}_{L}\nn \\
&=&[1\ 0]\ {\bf \Omega}
\l(\vec{V}_{R}-{{\bf \sigma_z}\ \vec{i}_{\rm bulk}\over {ge^2/h}}
+\ {i_{\rm back}\over {ge^2/h}}\l[\ba{r}1\\1\ea\r] \r)\nn \\
V_D&=&[0\ 1]\ {\bf \Omega}\ \vec{V}_{R}\nn \\
&=&[0\ 1]\ {\bf \Omega}
\l(\vec{V}_{L}+ {{\bf \sigma_z}\ \vec{i}_{\rm bulk}\over {ge^2/h}}
-\ {i_{\rm back}\over {ge^2/h}}\l[\ba{r}1\\1\ea\r] \r)
\label{bc1}
\eea
Hence
\bea
V_S-V_D&=&\vec{V}_R^T{\bf \Omega}\l[\ba{r}1\\-1\ea\r]
+{i_{\rm back}\over {ge^2/h}}-[1\ 0]
{{\bf \Omega}{\bf \sigma_z}\vec{i}_{\rm bulk}\over {ge^2/h}}\\
\label{bias1}
&=&\vec{V}_L^T{\bf \Omega}\l[\ba{r}1\\-1\ea\r]
+{i_{\rm back}\over {ge^2/h}}-[0\ 1]
{{\bf \Omega}{\bf \sigma_z}\vec{i}_{\rm bulk}\over {ge^2/h}}.
\label{bias2}
\eea
Let us first look at (\ref{bias1}).
Recall the decomposition (\ref{dec}).
Noticing that
${\bf \Omega}\ {\bf \sigma_z}\ {\bf \Omega}=g{\bf \sigma_z}$
is a diagonal matrix, one can rewrite the bias voltage
$V_S-V_D$ as
\be
V_S-V_D={h\over e^2}\l(I_R+{1\over g}\ i_{\rm back}-i_{+k_{\rm F}}\r),
\label{bias3}
\ee
i.e.,
$i_{-k_{\rm F}}{\bf \Omega}\l[\ba{r}0\\ 1\ea\r]$
in $\vec{i}_{\rm bulk}$ does not contribute to (\ref{bias3}).
In order to see the consequence of (\ref{bias3})
let us first consider the case: $i_{\rm bulk}=0$, i.e., electrons
are neither injected nor ejected through the bulk
contact.
In this case the total current in the presence
of the \rm backscattered current $i_{\rm back}$
is given by
$I=I_L=I_R={e^2\over h}(V_S-V_D)-{1\over g}\ i_{\rm back}$.
It is very interesting to compare it with
the current in the absence of {\it bulk} contact:
$I_{\rm max}={e^2\over h}(V_S-V_D)$.
The result is obviously
\be
I_{\rm max}-I={1\over g}\ i_{\rm back},
\label{1/g}
\ee
which is different from the naive expectation
$I_{\rm max}-I=i_{\rm back}$,
by a factor $1/g$.
In the derivation of (\ref{1/g})
the boundary condition (\ref{bc1})
has been treated carefully
with $V_S$ and $V_D$ being fixed.
Our result,  Eq. (\ref{1/g}), which differs from the analysis in
Ref. \cite{bena}, leads to an important remark on the shot-noise
experiment in TLL. \cite{future}

In order to see how the two-terminal conductance
is affected by the {\it bulk} contact, we connect
the {\it bulk} contact to the source (S) through an
ohmic resistance $r_S$.
In this case the total current circulating in the
system is $I=I_R=[1, -1]\ {\bf \Omega}\ \vec{V}_{R}$.
Therefore Eq. (\ref{bias3}) means
\be
I={e^2\over h}(V_S-V_D)-{1\over g}\ i_{\rm back}+i_{+k_{\rm F}}.
\label{source1}
\ee
Another remark is that in the physically
interesting case of small 
$i_{\rm bulk}=i_+ +i_-=i_{+k_{\rm F}}+i_{-k_{\rm F}}>0$,
$V=V_S-r_S i_{\rm bulk}$ where $r_S $ is a resistance between
the source and {\it bulk} contact. This means
$V_D\ll V < V_S$.

When the {\it bulk} contact is connected to the drain (D),
one should use (\ref{bias2}). Thus the total current
$I=I_L=[1, -1]\ {\bf \Omega}\ \vec{V}_{L}$
can be written as
\be
I={e^2\over h}(V_S-V_D)-{1\over g}\ i_{\rm back}-i_{-k_{\rm F}}.
\label{drain1}
\ee
We will be interested in the case of
small $i_{\rm bulk}<0$, i.e., $V_D< V=V_D-r_D i_{\rm bulk} \ll V_S$.

We will see in the next section that
Eqs. (\ref{source1},\ref{drain1})
give rise to a non-trivial physical consequence,
where we apply to them the results of our fractional scaling analysis.
This is not only due to the specific boundary condition
which we have chosen for the moment but also related to the existence
of a large stationary current.
In the absence of stationary current $I_{\rm bias}$, i.e.,
when $V_S=V_D=V_0$, which actually implies
$V_+=V_-=V_+^{(0)}=V_-^{(0)}=V_0$,
the scaling behavior of
$i_{\pm k_{\rm F}}$ is trivial:
$i_{+k_{\rm F}}=i_{-k_{\rm F}}
\sim\pm |V-V_0|^{2\Delta}$, depending on the
sign of $V-V_0$,
where $V$ is the electrostatic potential of the {\it bulk} contact
and $\Delta$ is the anomalous scaling dimension of a TLL
electron. In the case of finite SD voltage
$V_S\neq V_D$, where $V_+$, $V_-$, $V_+^{(0)}$ and $V_-^{(0)}$
are all different, we have to look into the details
of fractional decomposition and find out
the scaling law as a function of $I_{\rm bias}$.
This is done in Sec. V below.

\ss{One bulk contact with end contacts
in equilibrium with dressed eigenmodes}
Before turning to the RG analysis of {\it bulk} contacts,
let us briefly look at the second boundary condition
discussed in section III-B, i.e., the case where
the dressed eigenmode voltages are equilibrated with
those of the reservoirs: $V_L^{+}=V_{S}$, $V_R^{-}=V_{D}$.
The particles emitted from the left (right)
reservoir are moving to the right (left) and are in 
equilibrium with the
right (left)-moving eigenstates of the TLL.
This situation may be achieved in the
SWNT in the absence of screening gate.
The two-terminal conductance in the present case
in the {\it absence} of {\it bulk} contacts
is quantized at $G_{\rm SD}^{(2)}=ge^2/h$.
In the case of one {\it bulk} contact connected to
the source (S), the voltage drop equation
(\ref{drop2}) implies instead of (\ref{source1})
\be
I=g{e^2\over h}(V_S-V_D)-i_{\rm back}+i_{+},
\label{source2}
\ee
where the total current $I$ is given as
$I=I_R=g\ [1, -1]\ \vec{V}_{R}$.
When the {\it bulk} contact is connected to the drain (D), 
one finds instead
\be
I=g{e^2\over h}(V_S-V_D)-i_{\rm back}-i_{-}.
\label{drain2}
\ee
Note that in Eqs. (\ref{source2},\ref{drain2})
the correction due to the {\it bulk} contact enters
as either $i_{+}$ or $i_{-}$
in contrast to Eqs. (\ref{source1},\ref{drain1}).
As a result, we will see that the leading scaling exponents which appear
in Eqs. (\ref{source2},\ref{drain2}) are different from
the ones in (\ref{source1},\ref{drain1}) and that
they are simply related to the anomalous scaling dimension 
$\Delta$ of the TLL electron operator. All these issues
concerning the scaling behavior of Eqs.
(\ref{source1},\ref{drain1},\ref{source2},\ref{drain2})
will be discussed in detail in the next section.
However the reader who is more interested in the physical consequences
of the analysis in this section than a detailed deriviation
of RG equations can skip to Eqs. (\ref{scale1},\ref{scale2}) 
and Sec. V-C.

\section{RG analysis for the 3T conductance measurement}
We are mostly interested in
the leading order corrections to the conductance 
in the presence of {\it bulk} contact
under two different boundary conditions
discussed in Sec. III-A and Sec. III-B.
It is not difficult to realize
using the standard scaling analysis
that the backscattered current is IR-relevant 
whereas the bulk-injected current is UV-relevant
for repulsive interaction ($0<g<1$), i.e.,
$i_{\rm back}/(V_S-V_D)$ ($i_{\rm bulk}/(V_S-V_D)$)
scales to smaller (larger) values
as the SD voltage $V_S-V_D$ is increased.
However as we will see later in this section,
more precise understanding of the scaling will become
necessary in the analysis of Eqs.
(\ref{source1},\ref{drain1},\ref{source2},\ref{drain2}).
In particular it turns out less trivial in a chirally
separated system but essential in the discussion of conductance 
to identify the cutoff energy scale $\Lambda$
which defines the upper limit of {\it physically meaningful} 
energy scales.

In the case of
standard backscattering problem without $i_{\rm bulk}$, \cite{kane}
there are two fixed points on the RG flow diagram, usually drawn
in terms of the conductance $G_{\rm SD}^{(2)}$ ($y$-axis) 
as a function of $\Lambda$ ($x$-axis). The two fixed points are
at $G_{\rm SD}^{(2)}=0$ ($\Lambda=0$) and either at 
$G_{\rm SD}^{(2)}=e^2/h$ or at $G_{\rm SD}^{(2)}=ge^2/h$
($\Lambda\goto\infty$).
Here we are interested in the latter UV (ohmic) limit
in the presence of $i_{\rm bulk}$.
We will see that
the 3T SD conductance $G_{\rm SD}^{(3)}$
shows an UV-relevant deviation and interpolates between
the standard 2T values: $G_{\rm SD}^{(2)}=e^2/h$ 
and $G_{\rm SD}^{(2)}=ge^2/h$.

We start with the Euclidian Lagrangian density for non-chiral 
TLL,
\be
{\cal L}_{\rm TLL}=
{u\over 2}
\left[
{1\over g}
\l(\frac{\partial\Phi}{\partial x}\r)^2
+g\l(\frac{\partial\Theta}{\partial x}\r)^2
\right]
+i{\del\Phi\over\del\tau}
{\del\Theta\over\del x}.
\label{TLL}
\ee
We do not consider for the time being 
the effects of {\it end} contacts.
On the other hand we take into account
the existence of a stationary current $I_{\rm bias}$. 
In the absence of {\it bulk} contact,
the TLL has two chiral eigen modes respectively 
at voltages $V_+$ and $V_-$.
These voltages $V_+$ and $V_-$ are related to
the stationary current $I_{\rm bias}$ via
$I=g\ {e^2\over h}\ [1, -1]\ \vec{V}$.

The {\it bulk} contact is at $x=0$.
All the bosonic fields involved in \rm backscattering or tunneling
should be understood as those at $x=0$.
Therefore it is convenient to integrate out
the continuum degrees of freedom in (\ref{TLL})
to obtain the effective action at $x=0$.
\bea
S_0&=&{1\over\beta}\sum_\omega
|\omega|
\left(
{1\over g}
\l|\phi(\omega)\r|^2+g\l|\theta(\omega)\r|^2
\right)
\label{rep1}\\
&=&{g\over 2\beta}\sum_{\omega}
|\omega|
\left(
\l|\theta_+(\omega)\r|^2
+\l|\theta_-(\omega)\r|^2
\right),
\label{rep2}
\eea
where $\phi=\Phi(x=0),\theta=\Theta(x=0)$ and
$\theta_\pm=\theta\mp{\phi\over g}$.
In the \rm back-scattering problem,
\cite{kane}
$\theta$ is free and eventually can be integrated
out from the effective action, but let us for the moment keep $\theta$.
We need both fields in order to treat
1. \rm back-scattered current $i_{\rm back}$,
2. electron injection (ejection) $\vec{i}_{\rm bulk}$,
on the same footing. We will see that (\ref{rep2})
is the suitable way of writing $S_0$ for the 
latter problem.

\ss{Backward scattering}
We first consider the backscattering problem \cite{kane}.
The scattering potential due to the tunneling of quasiparticle reads
\bea
{\cal L}_{\rm back}
&\sim&\Gamma_{\rm back}\l[
\Psi_{+k_{\rm F}}^{\dagger}\Psi_{-k_{\rm F}}+\Psi_{+k_{\rm F}}\Psi_{-k_{\rm F}}^{\dagger}
\r]_{x=0}\nn \\
&\sim&\delta (x)\Gamma_{\rm back}\cos(2\sqrt{\pi}\phi),
\label{back}
\eea
i.e., $\theta$ is free and can be integrated
out from the effective action.
By throwing away $\theta$-part in (\ref{rep1});
the effective action at $x=0$ reduces to
\bea
S_{\rm back}(\Lambda)&=&{1\over\beta g}\sum_{|\omega|<\Lambda}
|\omega||\phi_\Lambda(\omega)|^2\nn \\
&+&\Gamma_{\rm back}(\Lambda) \int_0^\beta d\tau
\cos(2\sqrt{\pi}\phi_\Lambda).
\eea
This effective theory is meaningless unless we find
a suitable high-frequecy cutoff $\Lambda$, i.e.,
starting from bare cutoff $\Lambda_0$,
we integrate out the high-frequency unphysical
degrees of freedom down to $\Lambda$.
By decomposing the bosonic field $\phi$ in
Fourier space into fast ($\Lambda-d\Lambda<\omega<\Lambda$)
and slow ($0<\omega<\Lambda-d\Lambda$) parts and then
integrating out the high-frequency modes
$\Lambda-d\Lambda<\omega<\Lambda$, one obtains the RG
equation for $\Gamma_{\rm back}(\Lambda)$ as
$\Gamma_{\rm back}(\Lambda)=\Gamma_{\rm back}(\Lambda_0)
\l({\Lambda\over\Lambda_0}\r)^{g-1}$.
The scaling behavior of \rm backscattered current $i_{\rm back}$
is deduced from the RG equation by identifying
the physical $\Lambda$ to be
$\Lambda_{\rm back}=g\ e(V_+ -V_-)=V_S-V_D$ 
so that
$i_{\rm back}\sim (V_S-V_D)^{2g-1}$,
where we have chosen by convention as $V_S>V_D$ .
Notice that in the high-frequency (ohmic) limit ($V_S-V_D\goto \infty$)
the correction to the two-terminal conductance vanishes:
${di_{\rm back}\over d(V_S-V_D)}
\sim (V_S-V_D)^{2(g-1)}\goto 0$ as long as $0<g<1$.

\ss{Electron injection (ejection)}
For the tunneling of electron into or out of the
bulk sample, we basically follow the same spirit.
Through the {\it bulk} contact at $x=0$
electrons can tunnel into
the bulk sample which has two chirally separated eigen
modes respectively at voltages $V_+$ and $V_-$
from the electron reservoir at chemical potential $eV$.
An electron incident from
this electron reservoir at chemical potential $eV$
must be decomposed into two fractionally charged
quasiparticles in order to be absorbed in the bulk
sample.
The incident Fermi liquid electron
ends up in the final state with one of the
two possible electronic excitations of TLL,
i.e., either
$\Psi_{+k_{\rm F}}\sim
e^{i\sqrt{\pi}\ [1\ 0]\ {\bf \Omega}
\l[\ba{r}\theta_+\\ \theta_-\ea\r]}$
or
$\Psi_{-k_{\rm F}}\sim
e^{i\sqrt{\pi}\ [0\ 1]\ {\bf \Omega}
\l[\ba{r}\theta_+\\ \theta_-\ea\r]}$,
where $\Omega$ has been 
defined as
${\bf \Omega}={1\over 2}\l[\ba{rr}
1+g & 1-g \\
1-g & 1+g
\ea\r]$.
The scattering potential due to an injection or ejection
of electrons at $x=0$ is given, respectively, for the
$+k_{\rm F}$- and $-k_{\rm F}$-channel as
\bea
{\cal L}_{+k_{\rm F}}&=&\Gamma_{+k_{\rm F}}\l[
\Psi_{+k_{\rm F}}^\dagger\Psi_{g=1}
+\Psi_{+k_{\rm F}}\Psi_{g=1}^{\dagger}\r]\nn \\
&\sim&\Gamma_{+k_{\rm F}}\l[
\Psi_{g=1}^{\dagger}
e^{i\sqrt{\pi}\l({1+g\over 2}\theta_+ +{1-g\over 2}\ \theta_-\r)}
+({\rm h.c.})\r],
\label{+k_F}
\\
{\cal L}_{-k_{\rm F}}&=&\Gamma_{-k_{\rm F}}\l[
\Psi_{-k_{\rm F}}^\dagger\Psi_{g=1}
+\Psi_{-k_{\rm F}}\Psi_{g=1}^{\dagger}\r]\nn \\
&\sim&\Gamma_{-k_{\rm F}}\l[
\Psi_{g=1}^{\dagger}
e^{i\sqrt{\pi}\l({1-g\over 2}\theta_+ +{1+g\over 2}\ \theta_-\r)}
+({\rm h.c.})\r].
\label{-k_F}
\eea
The total effective action is
\bea
S_{\rm bulk}&=&{g\over 2\beta}
\left[
\sum_{|\omega|<\Lambda_+}|\omega|
\l|\theta_+(\omega)\r|^2+
\sum_{|\omega|<\Lambda_-}|\omega|
\l|\theta_-(\omega)\r|^2
\right]
\nn \\
&+&\int_0^\beta d\tau {\cal L}_{+k_{\rm F}}[\theta_+,\theta_-]
+\int_0^\beta d\tau {\cal L}_{-k_{\rm F}}[\theta_+,\theta_-],
\label{total}
\eea
where we have introduced two frequency cut-offs 
$\Lambda_+$ and $\Lambda_-$, respectively,
for $\theta_+$ and for $\theta_-$.
Notice that in Eq. (\ref{total}) 
$S_0$ is represented as (\ref{rep2}).
Now we derive the RG equation for 
$\Gamma_{+k_{\rm F}}(\Lambda_+,\Lambda_-)$
and $\Gamma_{-k_{\rm F}}(\Lambda_+,\Lambda_-)$.
using the standard perturbative RG analysis.
At leading order the two RG equations are decoupled
and can be treated independently.
Furthermore the cutoff frequencies
$\Lambda_+$ and $\Lambda_-$ can be different for
the two RG equations, since
available energy shells in $S_0$
for (\ref{+k_F}) and (\ref{-k_F}) can be differernt. 
The crucial step is, therefore, to find these cut-off
frequencies, i.e.,
$\Lambda_{+k_{\rm F}}^+$, $\Lambda_{+k_{\rm F}}^-$ 
for $\Gamma_{+k_{\rm F}}$, and
$\Lambda_{-k_{\rm F}}^+$, $\Lambda_{-k_{\rm F}}^-$ 
for $\Gamma_{-k_{\rm F}}$. 
In the Appendix these cutoffs are derived
based on a microscopic analysis where the fractionalization
of an incident electron is explicit.
By simply observing the physical processes in (\ref{+k_F}),
it is however not difficult to convince ourselves that 
the natural cutoffs for (\ref{+k_F}) are
$\Lambda_{+k_{\rm F}}^+=\Lambda_{+k_{\rm F}}^-=e\l(V-V_{+}^{(0)}\r)$,
where $V_{+}^{(0)}={1+g\over 2}eV_{+}+{1-g\over 2}eV_{-}$ is
the chemical potential of a $+k_{\rm F}$ electron
incident from two chiral eigenmodes of TLL.
Thus the RG equation for
$\Gamma_{+k_{\rm F}}$ reduces to that of a
single scaling parameter $\Lambda$,
i.e., following the same procedure as the
backscattering problem, one can find
the RG equation for $\Gamma_{+k_{\rm F}}(\Lambda)$ as
$\Gamma_{+k_{\rm F}}(\Lambda)=\Gamma_{+k_{\rm F}}(\Lambda_0)
\l({\Lambda\over\Lambda_0}\r)^{\Delta+1/2-1}$,
where
$\Delta={1\over 4}\l(g+{1\over g}\r)$
is nothing but the anomalous scaling dimension of 
a TLL electron operator.
The scaling behavior of bulk injected current $i_{+k_{\rm F}}$
is deduced from the RG equation by identifying
the physical $\Lambda$ to be 
$\Lambda=\Lambda_{+k_{\rm F}}^+=\Lambda_{+k_{\rm F}}^-
=e\l(V-V_{+}^{(0)}\r)$ as
\be
i_{+k_{\rm F}}\sim
\l\{\ba{l}
\l(V-V_{+}^{(0)}\r)^{2\Delta}
\ \ \ {\rm for}\ V>V_{+}^{(0)}
\\
-\l(V_{+}^{(0)}-V\r)^{2\Delta}
\ \ \ {\rm for}\ V<V_{+}^{(0)}
\ea\r..
\label{scale1}
\ee
Note that the energy dependence of Eq. (\ref{scale1})
is not a simple product of two chiral components
such as $(V-V_+)^\Delta (V-V_-)^\Delta$ as might be naively
expected.
The same argument applies for the
RG equation for $\Gamma_{-k_{\rm F}}$.
The scaling behavior of bulk injected current 
$i_{-k_{\rm F}}$ is obtained as
\be
i_{-k_{\rm F}}\sim
\l\{\ba{l}
\l(V-V_{-}^{(0)}\r)^{2\Delta}
\ \ \ {\rm for}\ V>V_{-}^{(0)}
\\
-\l(V_{-}^{(0)}-V\r)^{2\Delta}
\ \ \ {\rm for}\ V<V_{-}^{(0)}
\ea\r.,
\label{scale2}
\ee
where $eV_{-}^{(0)}={1-g\over 2}eV_{+}+{1+g\over 2}eV_{-}$
is the chemical potential of $-k_{\rm F}$ electron.
In the absence of stationary current $I_{\rm bias}$,
both Eqs. (\ref{scale1}) and (\ref{scale2})
reduce to the trivial scaling law:
$i_{+k_{\rm F}}=i_{-k_{\rm F}}
\sim\pm |V-V_0|^{2\Delta}$, depending on the
sign of $V-V_0$.

\ss{Conductances}
We now turn to the discussion of the physical consequences of 
Eqs. (\ref{source1},\ref{drain1}) and Eqs. (\ref{source2},\ref{drain2}),
which correspond, respectively, to the boundary conditions
discussed in Secs. III-A and III-B.
In Secs V-A and V-B, in particular,
in Eqs. (\ref{scale1}) and (\ref{scale2}),
we were able to relate the bulk injected
currents $i_{\pm k_{\rm F}}$ to local energy scales
in the vicinity of the {\it bulk} contact such as
$V-V_{+}^{(0)}$ or $V-V_{-}^{(0)}$.
In order to compare this term with other contributions
one has to rewrite them in terms of the source-drain
voltage $V_S-V_D$.
We focus on the case where the two {\it end} contacts
impose a large stationary current $I_{\rm bias}$ in the quantum wire
so that
\ben
\item
The chiral symmetry of the system should be preserved.
\item
The additional current injected through the {\it bulk} contact
can be treated as a small perturbation.
\een
We will see that
under the two boundary conditions discussed in Sec. III-A and
Sec. III-B,
the 3T SD conductance $G_{\rm SD}^{(3)}$
interpolates between the two {\it fixed points}:
$G_{\rm SD}=e^2/h$ and $G_{\rm SD}=ge^2/h$
in the ohmic limit.
We found non-trivial scaling exponents as a result of
specific boundary condition and electron fractionalization.

\sss{One bulk contact with end contacts in equilibrium with bare electrons}
In the case of equilibration with bare electrons
at the {\it end} contacts (see section III-A),
i.e., in the case of (\ref{source1},\ref{drain1}),
due to the boundary condition (\ref{bc1}),
$V_{+}^{(0)}$ in (\ref{scale1}) and $V_{-}^{(0)}$ in (\ref{scale2})
can be replaced, at leading order, by 
$V_{+}^{(0)}=V_S$ and $V_{-}^{(0)}=V_D$, respectively.
Now let us consider the case of Eq. (\ref{source1}),
in which the {\it bulk} contact is connected to the
source (S) through an ohmic resistance $r_S$.
We are mainly interested in the case:
$V_D\ll V=V_S-r_S i_{\rm bulk}<V_S$,
i.e., in the case of small $i_{\rm bulk}>0$,
since this gives the first correction to the
"end-contact model".
Substituting Eq. (\ref{scale1}) into Eq. (\ref{source1}), 
one can immediately see that the 
contribution to the total current $I$ of the bulk injected current obeys
a power law with an exponent $2\Delta$. 
The important result of our RG analysis, which is carefully derived in 
Sec. V-A and V-B as well as in the Appendix, is that this exponent appears as 
a power of the voltage difference $V_S-V$ and not of the source-drain voltage 
$V_S-V_D$. 

Now we have to rewrite this energy scale $V_S-V$ 
in terms of $V_S-V_D$. 
Let us notice that
$i_{+k_{\rm F}}\sim -(V_S-V)^{2\Delta}$ is negative
and very small when $V_D\ll V<V_S$
since $2\Delta={1\over 2}\l(g+{1\over g}\r)\ge 1$.
On the other hand,
$i_{-k_{\rm F}}\sim (V-V_{D})^{2\Delta}\sim (V_S-V_{D})^{2\Delta}$
gives a positive dominant contribution to
$i_{\rm bulk}$ compensating for the negative
$i_{+k_{\rm F}}$ to ensure
a positive $i_{\rm bulk}=i_{+k_{\rm F}}+i_{-k_{\rm F}}$.
Thus the small voltage difference
$V_S-V$ which appears in the scaling of
$i_{+k_{\rm F}}$ can be rewitten as a function of
the bias voltage $V_S-V_D$ as
$V_{S}-V\sim i_{\rm bulk}\sim i_{-k_{\rm F}}\sim (V_S-V_{D})^{2\Delta}$.
In other words, $V_S-V$ {\it itself} scales with the exponent $2\Delta$ 
in terms of the source-drain voltage $V_S-V_D$. 
The combination of these two $2\Delta$'s
leads to an unusual scaling behavior of  
the conductance measured in terms of
the bias voltage $V_S-V_D$:
\be
G_{\rm SD}^{(3)}={e^2\over h}-c_{1}(V_S-V_D)^{2(g-1)}
-c_{2}(V_{S}-V_D)^{(2\Delta)^2-1},
\label{result1}
\ee
where $c_1$ and $c_2$ are scale-invariant positive constants.
Eq. (\ref{result1}) constitutes one of the main
results of the paper,
since it not only contains a new type of
scaling behavior $(V_{S}-V_D)^{(2\Delta)^2-1}$
but also justifies our model
as a possible way to interpolate between the two
boundary conditions.
Notice that $(2\Delta)^2-1\ge 0$ for $0<g\le 1$.
Eq. (\ref{result1}) means that
even in the ohmic regime where the backscattering current
$i_{\rm back}$ scales to zero, the two-terminal conductance 
in the presence of the {\it bulk} contact
can be decreased 
due to the bulk-injected current $i_{\rm bulk}$
and interaction ($0<g<1$).

In the case of the {\it bulk} contact connected to the drain
(D), i.e., Eq. (\ref{drain1}), 
one finds, of course, the same universal behavior
as (\ref{result1})
as far as one focuses on the case $V_D< V \ll V_S$,
i.e., the leading order correction to the
boundary condition (\ref{bc0}).

\sss{One bulk contact with
end contacts in equilibrium with dressed eigenmodes}

Let us now consider the second boundary condition for
the {\it end} contacts: $V_S=V_R^+$, $V_D=V_L^-$,
i.e., we discuss the scaling behavior of
(\ref{source2},\ref{drain2}).
In this case, depending on whether the {\it bulk} contact
is connected to the source (S) or to the drain (D)
all the $V$'s in Eqs.
(\ref{cutoff1}),(\ref{cutoff2})
can be replaced either by $V=V_+$ or by $V=V_-$.
Then one can, of course, replace $V_+$ and $V_-$
respectively, by $V_S$ and by $V_D$.
Substituting $V=V_+=V_S$, $V_-=V_D$ into
Eqs. (\ref{cutoff1},\ref{cutoff2}), 
one finds $\Lambda_{+k_{\rm F}}^+=\Lambda_{+k_{\rm F}}^-=
{1-g\over 2}(V_{S}-V_{D})$ as well as
$\Lambda_{-k_{\rm F}}^+=\Lambda_{-k_{\rm F}}^-=
{1+g\over 2}(V_{S}-V_{D})$.
Eqs. (\ref{scale1}), (\ref{scale2}) may also be
rewritten accordingly.
Taking into account that 
$i_+={1+g\over 2}i_{+k_{\rm F}}+{1-g\over 2}i_{-k_{\rm F}}>0$
in contrast to the previous case:
$i_{+k_{\rm F}}<0$,
the leading scaling behavior of Eq. (\ref{source2})
is found to be
\be
G_{\rm SD}^{(3)}=g{e^2\over h}-c_{1}(V_S-V_D)^{2(g-1)}
+c_{3}(V_{S}-V_D)^{2\Delta-1},
\label{result2}
\ee
where $c_1$ and $c_3$ are scale-invariant positive constants.
Comparing with (\ref{result1}),
one can observe that in Eq. (\ref{result2}),
\ben\item
The existence of the bulk current
increases $G_{\rm SD}^{(3)}$, indicating that 
the correction indeed
interpolates between the two limiting cases, i.e.,
from $G_{\rm SD}=ge^2/h$ to $G_{\rm SD}=e^2/h$.
\item
The correction due to the bulk current
does not scale in the same way.
Note that 
the exponent of Eq. (\ref{result2}) is
simply related to the anomalous scaling dimension of
a TLL electron operator.
\een

\sss{Remarks on energy scales}
Before ending this section, we would like to make
some remarks on the energy scales where
Eqs. (\ref{result1},\ref{result2}) are valid.
\ben\item
We have neglected the backscattering at the interface
between the 1D sample and the reservoirs.
This is justified when $e(V_S-V_D)\gg \Lambda_{\rm end}$,
where $\Lambda_{\rm end}$ is a crossover energy
associated with the
quasiparticle$\between$electron tunneling duality model
at effective filling factor $\nu_{\rm eff}$
\cite{chamon,fn3}.
\item
In order for Eqs. (\ref{result1},\ref{result2}) to be valid
it is also required that backscatterings
in the bulk (at the {\it bulk} contact or due to some impurities
in the bulk) can be treated perturbatively.
This situation is achieved when $e(V_S-V_D)\gg \Lambda_{\rm back}$,
where $\Lambda_{\rm back}$ is another crossover energy scale:
$\Lambda_{\rm back}\propto\Gamma_{\rm back}(\Lambda_0)^{1/(1-g)}$.
\cite{kane}
\item
Finally $V_S-V_D$ must be sufficiently large so that
the perturbative analysis for the electron injection (ejection)
can be justified, i.e.,
$e(V_S-V_D)\gg \Lambda_{\rm bulk}$, where
$\Lambda_{\rm bulk}\propto\Gamma_{\pm k_F}(\Lambda_0)^{2/(1-2\Delta)}$,
$\Delta={1\over 4}\l(g+{1\over g}\r)$.
\een
We have also assumed that
\ben\item
the temperature $T$ is sufficiently low,
\item
the typical size $L$ of the system is large enough,
\een
so that $e(V_S-V_D)\gg T, u/L$
should be satisfied,
where $u=v_F/g$ is the velocity of the chiral eigen modes.

\vspace{0.2cm}
We studied the effects of one {\it bulk} contact 
as a leading order correction to the two limiting cases of
{\it end contact model}: 
(A) $G_{\rm SD}=e^2/h$ and (B) $G_{\rm SD}=ge^2/h$,
corresponding, respectively, to the
(A) the presence and (B) the absence of screening
by a metallic gate discussed in Sec. III. 
We found,
\ben\item
In both cases
$G_{\rm SD}^{(3)}$ is not quantized even
in the ohmic limit $V_S-V_D\goto\infty$,
interpolating between the two limiting cases:
(A) $G_{\rm SD}=e^2/h$ and (B) $G_{\rm SD}=ge^2/h$.
\item
The correction due to the bulk current, however,
does not scale in the same way
(see Eqs. (\ref{result1}) and (\ref{result2})).
In the presence of screening metallic gate,
i.e., in the case of Eq. (\ref{result1}),
it exhibits a pronounced scaling behavior:
$(V_S-V_D)^{(2\Delta)^2-1}$, where
$2\Delta={1\over 2}\l(g+{1\over g}\r)>1$ ($2\Delta=1$) for (non)
interacting case.
\een
In terms of the RG picture introduced at the beginning of this
section, UV-relevant deviation from the standard
two fixed points: 
(A) $G_{\rm SD}=e^2/h$ and (B) $G_{\rm SD}=ge^2/h$
may suggest a possibility of new continuous fixed points
between the above two values.

\sec{Generalization to many bulk contacts}
In this section we continue the analysis
of the {\it bulk} contacts.
Some of the electrons are injected into the nanotube
through $N_L$ {\it bulk} contacts in the left reservoir
(others are through the {\it end} contact).
Both of them contribute to the total current $I$ which flows
through the sample.
Similarly in the right reservoir
some electrons are ejected from the nanotube
not only at both ends but also through one of the $N_R$
{\it bulk} contacts.
In brief we generalize the analysis of {\it one bulk} contact
to {\it many bulk} contacts.

This model is clearly inspired by the work of Chamon and Fradkin. 
The curious result derived in Ref. \cite{chamon} is 
that the conductance is not monotonous 
as a function of $N_R$ and $N_L$ (See Sec. III-C), 
and has a sort of damped oscillatory behaviour 
which depends on the parities of $N_R$ and $N_L$.
\cite{chiral}
So we were curious to determine 
if such a non-monotonous behaviour would 
also be predicted in the case of the non-chiral
TLL, and if the renormalized 
conductance could be experimentally measured in the 
case of many contacts,
as opposed to the situation found in \cite{safi1,maslov,pono1}.
As will appear below, our answer
is that the non-chiral TLL 
does not behave like a chiral one as far as
the number of  contacts is concerned, although the 
conductance results
are affected when weak contacts are applied 
in the bulk of the sample
strongly connected to reservoir through its end points.

Another motivation behind this type of model 
is the experiment by Kasumov et. al.
\cite{kasumov},
where they found a clear signature
of superconduting behavior in isolated SWNT samples
as well as nanotube ropes.
In this experiment
the isolated SWNT sample $ST_1$ exhibits a resistance $R$
which saturates at high temperatures to $R\sim 25 {\rm k\Omega}$
(see Fig. 2B of Ref. \cite{kasumov}).
As has been already mentioned,
this corresponds to the conductance
$G_{\rm SD}\sim e^2/h$,
which is smaller by a factor $1/4$
than $G_{\rm SD}=4e^2/h$ expected from the boundary condition
(\ref{bc0}). 
In this experiment
the nanotubes are embedded (melt) into the reservoirs 
at both ends.
Given the finite radius of the nanotube ($\sim 1.5 {\rm nm}$)
and possible roughness
of the electrode surface from which the nanotube emerges,
it is legitimate to question the validity of 
single point contact model.
What is hoped here is that our model with a weak bulk
contact is a first step towards a proper description of this
experimental situation.

\ss{Many bulk contacts with end contacts in equilibrium
with bare electrons}
Consider $N_R+N_L$ independent {\it bulk} contacts
$N_L$ of which are connected to the source (S) 
and the rest of which to the drain (D).
The boundary condition (\ref{bc1})
is generalized to
\be
\l\{\ba{l}
V_S=V_{-N_L}^{+(0)}
=[1\ 0]\ {\bf \Omega}\ \vec{V}_{-N_R}\nn \\
V_D=V_{N_R}^{-(0)}
=[0\ 1]\ {\bf \Omega}\ \vec{V}_{N_L}
\ea\r.
\label{bcmany}
\ee

\vspace{-2.0cm}
\begin{figure}[h]
\epsfig{figure=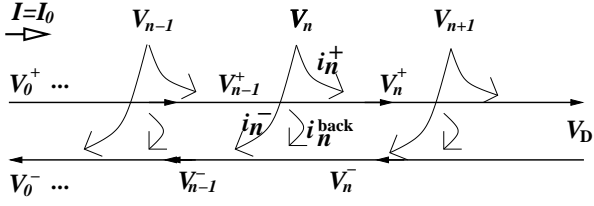,width=6.5cm,angle=-90}
\vspace{-2.0cm}
\caption{Many bulk contacts.}
\end{figure}

Let us focus on the $n$th {\it bulk} contact ($n=1,\cdots,N_R$),
which is, by definition, connected to the drain (D).
The voltage equations at $n$th {\it bulk} contact
can be written as
$i_{n}^{+}-i_{n}^{\rm back}=
g{e^2\over h}(V_{n-1}^{+}-V_{n}^+)$,
$i_{n}^{-}+i_{n}^{\rm back}=
g{e^2\over h}(V_{n}^{-}-V_{n-1}^-)$.
$i_{n}^+$ ($i_{n}^-$) is a current {\it injected}
into the TLL eingenmode with $+$($-$)-chirality.
The definition of eigenmode voltages
$V_{n}^{\pm}$ are given in Fig. 2.
Using the vector notation
$\vec{i}_n=\l[\ba{r}
i_{n}^+\\i_{n}^-\ea\r]$,
$\vec{V}_n=\l[\ba{r}
V_{n}^+\\V_{n}^-\ea\r]$,
one may rewrite the voltage drop equation as
$\vec{V}_{n-1}-\vec{V}_{n}
={{\bf \sigma_z}\vec{i}_n^{\ \rm bulk}\over ge^2/h}
-{i_n^{\rm back}\over ge^2/h}\l[\ba{r}1\\1\ea\r]$.
The total current circulating in the system is
defined as
$I=I_0=[1, -1]{\bf \Omega}\vec{V}_0$.
Using recursively the voltage equation, one finds
$V_{N_R}=\vec{V}_0+
\sum_{n=1}^{N_R}\l(
{{\bf \sigma_z}\vec{i}^{\ \rm bulk}_{n}\over ge^2/h}-
{i^{\rm back}_{n}\over ge^2/h}
\l[\ba{r}1\\1\ea\r]\r)$.
Following the same procedure,
one will find
a similar relation for the left contact:
$V_{-N_L}=\vec{V}_0-
\sum_{n=1}^{N_L}\l(
{{\bf \sigma_z}\vec{i}^{\ \rm bulk}_{-n}\over ge^2/h}-
{i^{\rm back}_{-n}\over ge^2/h}
\l[\ba{r}1\\1\ea\r]\r)$.

Using recursively these voltage drop equations, one finds
\bea
V_S&=&[1\ 0]{\bf \Omega}
\l[\vec{V}_0-
\sum_{n=1}^{N_L}\l(
{{\bf \sigma_z}\vec{i}^{\rm\ \rm bulk}_{-n}\over ge^2/h}-
{i^{\rm back}_{-n}\over ge^2/h}
\l[\ba{r}1\\1\ea\r]\r)\r],
\nn \\
V_D&=&[0\ 1]{\bf \Omega}
\l[\vec{V}_0+
\sum_{n=1}^{N_R}\l(
{{\bf \sigma_z}\vec{i}^{\rm\ \rm bulk}_{n}\over ge^2/h}-
{i^{\rm back}_{n}\over ge^2/h}
\l[\ba{r}1\\1\ea\r]\r)\r].
\nn
\eea
Thus the bias voltage $V_S-V_D$ can be expressed
in terms of $\vec{V}_0$, $i^{\rm back}_{\pm n}$
and $\vec{i}_{\pm n}$ as
\bea
V_S-V_D=
[1, -1]{\bf \Omega}\vec{V}_0
&+&
\sum_{n=1}^{N_L}{i^{\rm back}_{-n}\over ge^2/h}+
\sum_{n=1}^{N_R}{i^{\rm back}_{n}\over ge^2/h}
\nn \\
-[1\ 0]{\bf \Omega}{\bf \sigma_z}\sum_{n=1}^{N_L}
{\vec{i}^{\ \rm bulk}_{-n}\over ge^2/h}
&-&[0\ 1]{\bf \Omega}{\bf \sigma_z}\sum_{n=1}^{N_R}
{\vec{i}^{\ \rm bulk}_{n}\over ge^2/h}
\label{many}
\eea
Recall that 
${\bf \Omega\sigma_z\Omega}=g{\bf \sigma_z}$.
Using the decomposition analogous to (\ref{dec}),
one can see that
$i_{-n}^{-k_{\rm F}}$ ($n=1,\cdots,N_L$) and
$i_{n}^{+k_{\rm F}}$ ($n=1,\cdots,N_R$)
do not contribute to (\ref{many}).
The total current circulating in the system is
$I=I_0=[1, -1]{\bf \Omega}\vec{V}_0$.
Since we are interested in the first correction
to (\ref{bc0}), we consider the case where
all the {\it bulk} contacts connected to the source (S)
are at voltages $V_D<V \ll V_S$ and
all the {\it bulk} contacts connected to the drain (D)
are at voltages $V_D\ll V<V_S$.
If this is the case,
the leading scaling behavior of the conductance
in terms of the bias voltage $V_S-V_D$
obtained from Eq. (\ref{many})
reduces to (\ref{result1}) independently
of $N_R$ and $N_L$.
Eq. (\ref{result1}), therefore, the non-chiral version
of \cite{chiral} does not exhibit
an oscillatory behavior as a function of
$(N_R,N_L)$. 
Nevertheless Eq. (\ref{result1})
indeed interpolates between the two boundary conditions
discussed in Sec. III, which was also the case in
\cite{chiral}.

\ss{Many bulk contacts with end contacts
in equilibrium with dressed eigenmodes}
In the case of equilibration with dressed eigenmodes
the boundary condition (\ref{bcmany})
should be replaced by
$V_S=V_{-N_R}^{+}$, $V_D=V_{N_L}^{-}$.
Using the same
voltage drop equations, one can easily see,
using the decomposition analogous to (\ref{dec}),
that $i_{-n}^{-}$ ($n=1,\cdots,N_R$) and
$i_{n}^{+}$ ($n=1,\cdots,N_L$)
do not contribute to 
the total current
$I=I_0=g{e^2\over h}[1, -1]\vec{V}_0$.
Then the leading scaling behavior of the conductance
in terms of the bias voltage
reduces to (\ref{result2}) independently
of $N_R$ and $N_L$.

\sec{TLL with internal degrees of freedom --- Application to SWNT and MWNT}

Up to now we have considered for simplicity
spinless TLL model, where 
$2\Delta={1\over 2}\l(g+{1\over g}\r)$.
In order to apply the above results, in particular,
Eqs. (\ref{result1}) and (\ref{result2}) for nanotubes
let us recall the following properties of
SWNT and MWNT:
The SWNT have four conducting
channels: two subbands $\times$ (charge, spin)
at room-temperatures, 
indicating an expected quantized conductance: $G_{\rm SD}=4e^2/h$.

The experimental data for conductance measurements
in SWNT and MWNT display a variety of results.
The ballistic transport in carbon nanotubes 
was first observed in MWNT, showing the conductance
$G_{\rm SD}\sim 2e^2/h$
\cite{deheer}.
This implies that in MWNT only one of the two
subband contributes to the electronic transport.
In the case of SWNT,
the observed conductance exhibits fluctuations versus Fermi energy
approaching to the theoretically expected value:
$G_{\rm SD}=4e^2/h$ as the temperature is lowered.
\cite{laughlin}

Let us now focus on the case of SWNT. The spinless TLL theory
studied in earlier sections should be generalized
to acquire $2\times 2=4$ flavors $f=c+, c-, s+, s-$.
The four channels are obtained from combining charge
($c$) and spin ($s$) degrees of freedom as well as
symmetric  ($+$) and anti-symmetric ($-$)
linear combinations of the two Fermi points.
Correspondingly we must introduce
four TLL parameters: $g_{c+}, g_{c-}, g_{s+}, g_{s-}$.
The scaling dimension $\Delta$ of TLL electron operator
can be written, for example, 
in terms of these TLL parameters as
$\Delta_{\rm SWNT}={1\over 16}
\sum_{f}\l(g_f+{1\over g_f}\r)$.
Whereas the charge conductance $G_{\rm SD}$ is determined 
only by $g_{c+}$, i.e.,
$G_{\rm SD}=4g_{c+}e^2/h$, as was the case for TLL
with spin \cite{furusaki}.
In any case one can verify by carefully investigating the
effective Coulomb interaction in SWNT
\cite{egger1}
that the interaction gives rise to
a significant renormalization
only for $g_{c+}$, whereas
$g_f\sim 1$ for $f=c-, s+, s-$ (neutral modes).
Thus the TLL parameter $g$ for SWNT is defined as
$g=g_{c+}$, which is estimated to be typically
in the range $0.2-0.3$.

To summarize one has to make the following
replacements 
in order to apply Eqs. (\ref{result1}) and (\ref{result2}) 
for SWNT:
\ben\item
The anomalous scaling dimension $\Delta$ of TLL electron operator
should be replaced by 
$\Delta_{\rm SWNT}={1\over 16}\l(g+{1\over g}\r)+{3\over 8}$.
\item
The ohmic conductance
in equilibrium either with bare electrons ($G_{\rm SD}^{(2)}=e^2/h$)
or with dressed eigen modes ($G_{\rm SD}^{(2)}=ge^2/h$) 
should be multiplied by 4, in order to accont for the
number of conducting channels.
\een
Apart from these changes, however,
the main claims of the preceding sections remain unchanged.

\sec{Discussion and Conclusions}
In the first half of this paper we argued that in the case of
standard end-contact geometry,
the two-terminal conductance $G_{\rm SD}^{(2)}$ in the
ohmic limit can be either 
$G_{\rm SD}^{(2)}=e^2/h$ or $G_{\rm SD}^{(2)}=ge^2/h$ 
depending on the boundary conditions.
In our point of view, different boundary conditions apply
in the presence or absence of a metallic gate 
close to the 1D sample.

It is plausible that the boundary conditions
studied in refs. \cite{safi1,maslov,pono1,alek2,safi2,egger2}
are realized when a gate is present close enough to
the wire. This is suggested by the agreement between
the result of Ref. \cite{tarucha}
and the theoretical analysis using boundary conditions
such that bare reservoir electrons are not in
equilibrium with dressed eigenmode of the TLL,
but with bare particles inside the TLL.

On the other hand, more recent experiments suggest
that a variety of other boundary conditions
are realized experimentally
\cite{yacoby,picciotto1,picciotto2,kasumov}.
It is tempting to interpret
the result of Kasumov et. al. 
\cite{kasumov} as a consequence of
boundary conditions such that bare reservoir electrons
are in equilibrium with dressed TLL eigenmodes.

The absence of a gate in this experiment
suggests that long-range interactions inside the
carbon nanotube are instrumental in bringing about this
different boundary condition.
If our analysis is correct, a check would be
to measure the nanotube conductance in the presence
of a metallic gate sufficiently
close to the nanotube compared with the sample length
for the interactions to be screened.
Then we would expect $G_{\rm SD}^{(2)}=4e^2/h$.
It is striking that the result of Kasumov et. al.,
if interpreted as $G_{\rm SD}=4ge^2/h$, where $g$ would
be the TLL interaction parameter, yields a value
$g\sim 0.25$ in very good agreement with the
theoretical value calculated in Ref.
\cite{egger1}.
Other experiments 
\cite{yacoby,picciotto1,picciotto2}
clearly suggest other boundary conditions
\cite{future}.

Now the nature of two fixed points was understood
as different boundary conditions at the {\it end} contacts
by making clear distinction
between the {\it bare} and {\it dressed eigenmode} densities 
in the bosonized formulation.
In the second half of the paper we proposed a system of 1D sample
coupled to  {\it \rm bulk} contacts as well as
{\it end} contacts where we found qualitatively
different behaviors of the conductance,
e.g., different scaling dimensions,
as a consequence of a large stationary current.
As a result, we found that the addition of {\it bulk} contacts
interpolates between the two fixed points.
The RG analysis for this problem has been
developped by taking into account
explicitly the {\it fractionalization} of 
electronic charge. 

We studied in particular the leading scaling behavior of the
corrections to the two-terminal conductance $G_{\rm SD}^{(3)}$
in the presence of bulk-injected current
in (A) the presence and (B) the absence of screening
by a metallic gate. 
We found in both cases
$G_{\rm SD}^{(3)}$ is not quantized even
in the ohmic limit $V_S-V_D\goto\infty$,
interpolating between the two limiting cases:
(A) $G_{\rm SD}=e^2/h$ and (B) $G_{\rm SD}=ge^2/h$.
The correction due to the bulk current, however,
does not scale in the same way
(see Eqs. (\ref{result1}) and (\ref{result2})).
In the case of Eq. (\ref{result1}), corresponding to
the equilibration with {\it bare} electrons (Sec. III-A),
it exhibits, as a consequence of this particular
boundary condition (\ref{bc0})) or more precisely Eq. (\ref{bc1})),
a pronounced scaling behavior:
$(V_S-V_D)^{(2\Delta)^2-1}$, where
$2\Delta={1\over 2}\l(g+{1\over g}\r)$.
The understanding of strong coupling limit for the {\it bulk} contact
was left for future study. 
The discussion on the shot-noise spectrum under a variety of
boundary conditions discussed in this paper is
obviously of interest. This will be discussed in a
forthcoming publication
\cite{future}.

\ack
We are grateful to H\'el\`ene Bouchiat and 
In\`es Safi for useful discussions.
K. I. is supported by JSPS Postdoctoral Fellowships for
Research Abroad.

\begin{appendix}
\sec{RG analysis for fractional particles}
We derive the RG equation for 
$\Gamma_{+k_{\rm F}}(\Lambda_+,\Lambda_-)$
and $\Gamma_{-k_{\rm F}}(\Lambda_+,\Lambda_-)$
starting from the effective action
(\ref{total}) with (23,24). 
At leading order the two RG equations are decoupled
and can be treated independently.
The cutoff frequencies
$\Lambda_+$ and $\Lambda_-$ can be different for
the two RG equations, since
available energy shells in $S_0$
for (\ref{+k_F}) and (\ref{-k_F}) can be different. 
The crucial step was, therefore, to find these cut-off
frequencies, i.e.,
$\Lambda_{+k_{\rm F}}^+$, $\Lambda_{+k_{\rm F}}^-$ 
for $\Gamma_{+k_{\rm F}}$, and
$\Lambda_{-k_{\rm F}}^+$, $\Lambda_{-k_{\rm F}}^-$ 
for $\Gamma_{-k_{\rm F}}$. 

When an electron incident from
the electron reservoir at chemical potential $eV$
tunnels into the TLL,
it must be decomposed into two fractionally charged
quasiparticles in order to be absorbed in the bulk
sample.
This final state TLL electron has either of
the following energies
$eV_{+}^{(0)}=[1\ 0]{\bf \Omega}\l[\ba{r}eV_+\\ eV_-\ea\r]$,
$eV_{-}^{(0)}=[0\ 1]{\bf \Omega}\l[\ba{r}eV_+\\ eV_-\ea\r]$,
respectively, 
for $\Gamma_{+k_{\rm F}}$ and for $\Gamma_{-k_{\rm F}}$. 
In contrast
the energy decomposition of the incident Fermi liquid 
electron is quite arbitrary,
i.e.,
$eV=[1\ 0]{\bf \Omega}\l[\ba{r}eV'_+\\ eV'_-\ea\r]$
for $\Gamma_{+k_{\rm F}}$
and
$eV=[0\ 1]{\bf \Omega}\l[\ba{r}eV'_+\\ eV'_-\ea\r]$
for $\Gamma_{-k_{\rm F}}$, 
where $V'_+$ ($V'_-$) is a part of the
electrostatic potential attributed to  
the $+$($-$)-chirality.
The only constraint is that both $V'_+-V_+$ and $V'_- -V_-$
should have the same sign, i.e.,
they are positive (negative) when the current is injected
(ejected).
Taking into account this constraint, one can count
the available energy shells for tunneling.
These procedures are schematically explained in
Fig. 3.

\vspace{-2.0cm}
\begin{figure}[h]
\epsfig{figure=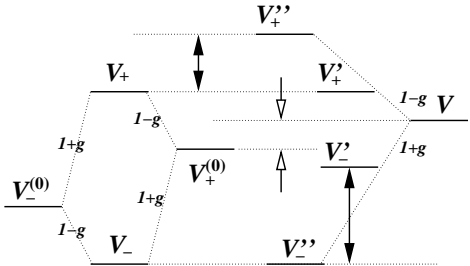,width=6.5cm,angle=-90}
\vspace{-1.0cm}
\caption{Available energy shells for the tunneling into TLL.}
\end{figure}

Let us focus on the RG equation for $\Gamma_{+k_{\rm F}}$,
i.e., for the tunneling of $+k_{\rm F}$ electron.
On the left it is shown that the final state TLL electron 
has either of the following energies 
$eV_{+}^{(0)}=[1\ 0]{\bf \Omega}\l[\ba{r}V_+ \\ V_-\ea \r]$, 
$eV_{-}^{(0)}=[0\ 1]{\bf \Omega}\l[\ba{r}V_+ \\ V_-\ea \r]$, respectively, 
for $\Gamma_{+k_{\rm F}}$ and for $\Gamma_{-k_{\rm F}}$.
On the right we focus on the tunneling of $+k_{\rm F}$ electron incident 
from the {\it bulk} contact at chemical potential $eV$. 
We consider the case 
$V_+^{(0)}<V<V_+$. 
The energy of the incident elecron is decomposed into 
each chirality as 
$eV=[1\ 0]{\bf \Omega}\l[\ba{r}eV'_+\\ eV'_-\ea\r]$. 
In order for a current to be injected, $V'_+\ge V_+$ must be satisfied. 
When $V'_+=V_+$, $V-V'_-={1+g\over 1-g}(V_+ -V)$. 
Starting from this value, 
$V'_-$ can take values down to $V_-$. 
When $V'_-$ reaches this limit, 
where we redefine the energy decomposition of
the incident electron as 
$eV=[1\ 0]{\bf \Omega}\l[\ba{r}eV''_+\\ eV''_-\ea\r]$, 
therefore, $V''_-=V_-$, the other $V''_+$ satisfies 
$V''_+ -V={1-g\over 1+g}(V-V_-)$. 
The cutoff energy scales which appear in the RG equation for 
$\Gamma_{+k_{\rm F}}$ are determined as 
$\Lambda_{+k_{\rm F}}^+={1+g\over 2}e(V''_+ -V_+)$, 
$\Lambda_{+k_{\rm F}}^-={1-g\over 2}e(V'_- -V_-)$. 

Thus we were able to derive on microscopic grounds the energy cutoffs which 
we have used to find (26),
\bea
\Lambda_{+k_{\rm F}}^+&=&{1+g\over 2}e 
\l[{1-g\over 1+g}(V-V_-)-(V_+-V)\r]\nn \\
&=&e\l(V-V_{+}^{(0)}\r),
\nn \\
\Lambda_{+k_{\rm F}}^-&=&{1-g\over 2}e 
\l[(V-V_-)-{1+g\over 1-g}(V_+-V)\r]\nn \\
&=&e\l(V-V_{+}^{(0)}\r),
\label{cutoff1}
\eea
i.e., $\Lambda_{+k_{\rm F}} ^+=\Lambda_{+k_{\rm F}} ^-$.
Once these energy scales are determined,
one can employ the RG analysis for $\Gamma_{+k_{\rm F}}$..

The same argument applies for the
RG equation for $\Gamma_{-k_{\rm F}}$.
The cutoff frequencies for the tunneling of $-k_{\rm F}$ electron
are obtained as
\bea
\Lambda_{-k_{\rm F}}^+&=&{1-g\over 2}e 
\l[{1+g\over 1-g}(V-V_-)-(V_+-V)\r]\nn \\
&=&e\l(V-V_{-}^{(0)}\r)
\nn \\
\Lambda_{-k_{\rm F}}^-&=&{1+g\over 2}e 
\l[(V-V_-)-{1-g\over 1+g}(V_+-V)\r]\nn \\
&=&e\l(V-V_{-}^{(0)}\r)
\label{cutoff2}
\eea
Hence the RG equation for $\Gamma_{-k_{\rm F}}$
takes the same form as that of $\Gamma_{+k_{\rm F}}$
except that one should identify
$\Lambda$ to be
$\Lambda=\Lambda_{-k_{\rm F}}^+=\Lambda_{-k_{\rm F}}^-=
V-V_{-}^{(0)}$.
\end{appendix}

\refs

\bibitem{pham} 
K.-V. Pham, M. Gabay and P. Lederer,
Phys. Rev. B {\bf 61}, 16397 (2000).

\bibitem{goldman}
V. J. Goldman and B. Su, Science {\bf 267}, 1010-1012 (1995);  
V. J. Goldman, I. Karakurt, Jun Liu and A. Zaslavsky, 
Phys. Rev. {\bf B 64}, 85319 (2001).

\bibitem{noise} 
L. Saminadayar, D. C. Glattli, Y. Jin, B. Etienne, 
Phys. Rev. Lett. {\bf 79}, 2526 (1997). 
R. de-Piccitio, M. Reznikov, M. Heiblum, V. Umansky, G. Bunin 
and D. Mahalu, Nature {\bf 389}, 162, (1997).

\bibitem{hal}
F.D.M. Haldane, Phys. Rev. Lett., {\bf 45}, 1358, (1980);
ibid., J. Phys. C{\bf 14}, 2585, (1981).

\bibitem{safi1}
I. Safi and H.J. Shulz,
ibid., R17040 (1995).
See also I. Safi and H.J. Shulz, in 
{\it Quantum Transport in Semiconductor
Submicron Structures}, ed. by B. Kramer
(Kluwer Academic Press, Dordrecht, 1995);

\bibitem{maslov}
D.L. Maslov and M. Stone,
Phys Rev. Lett. {\bf B52},  R5539 (1995).

\bibitem{pono1}
V.V. Ponomarenko,
Phys Rev. Lett. {\bf B52}, R8666 (1995).

\bibitem{apel}
W. Apel and T.M. Rice, Phys Rev. B {\bf 26}, 7063 (1982).

\bibitem{tarucha}
S. Tarucha, T. Honda and T. Saku, Solid State Commun.
{\bf 94}, 413 (1995).

\bibitem{wen}
X.G. Wen, Phys. Rev. B {\bf 41}, 12838
(1990); Adv. Phys. {\bf 44}, 405 (1995)

\bibitem{tsui}
D.C. Tsui, H.L. Stormer and A.C. Gossard,
Phys Rev. Lett. {\bf 48}, 1559 (1982).

\bibitem{yacoby}
A. Yacoby, H.L. Stormer, N.S. Wingreen, L.N. Pfeiffer,
K.W. Baldwin and K.W. West,
Phys. Rev. Lett. {\bf 77}, 4612 (1996).

\bibitem{picciotto1}
R. de Picciotto, H.L. Stormer, A. Yacoby, L.N. Pfeiffer,
K.W. Baldwin and K.W. West,
Phys. Rev. Lett. {\bf 85} 1730 (2000).

\bibitem{picciotto2}
R. de Picciotto, H.L. Stormer, L.N. Pfeiffer,
K.W. Baldwin and K.W. West,
Nature {\bf 411} 51 (2001).

\bibitem{bockrath}
M. Bockrath, D.H. Cobden, J. Lu, A.R. Rinzler,
R.E. Smalley, L. Balents and P.L. McEuen,
Nature, {\bf 397}, 598 (1999).

\bibitem{laughlin}
J. Kong, E.Y. Yenilmez, T.W. Tombler, W. Kim, H. Dai,
R.B. Laughlin, L. Liu, C.S. Jayanthi and S.Y. Wu, 
Phys Rev. Lett. {\bf 87}, 106801 (2000).

\bibitem{kasumov}
A.Yu. Kasumov, R. Deblock, M. Kociak, B. Reulet,
H. Bouciat, I.I. Khodos, Yu.B. Gorbatov, 
V.T. Volkov, C. Journet and M. Burghard,
Science, {\bf 284}, 1508 (1999).

\bibitem{saito}
R. Saito, G. Dresselhaus and M.S. Dresselhaus,
{\it Physical Properties of Carbon Nanotubes},
Imperial College Press (1998).

\bibitem{dekker}
C. Dekker, Phys. Today, {\bf 52}, No. 5, 22 (May 1999).

\bibitem{iijima}
S. Iijima, Nature, {\bf 354}, 56 (1991).

\bibitem{egger1}
R. Egger, A. Bachtold, M. S. Fuhrer,
M. Bockrath, D. H. Cobden, and P. L. McEuen,
cond-mat/0008008.

\bibitem{deheer}
S. Frank, P. Poncharal, W.A. de Heer,
Science, {\bf 280}, 1744 (1998).

\bibitem{gogolin}
R. Egger and A.O. Gogolin,
Phys Rev. Lett. {\bf 79}, 5082 (1997);
C.L. Kane, L. Balents and M.P.A. Fisher,
ibid., 5086 (1997).

\bibitem{landauer}
R. Landauer, Phil. Mag. {\bf 21}, 863 (1970);
M. B\"uttiker, Phys. Rev. {\bf B38} 9375 (1988).

\bibitem{chamon}
C. Chamon and E. Fradkin, Phys. Rev. {\bf B56}, 2012 (1997).

\bibitem{future}
In this work we shall consider only the above
two boundary conditions as far as {\it end} contacts
are concerned ---
A more general thermodynamic approach including
a prediction on the shot-noise spectrum under a variety of
boundary conditions will be discussed in
K.-I. Imura, K.-V. Pham, P. Lederer and F. Pi\'echon,
in preparation.

\bibitem{fn1}
Of course, the stationary component of
$\rho^{\rm (0)}_{\pm}(x,t)$ corresponds to the
number of {\it bare} excitations, i.e.,
those at $+k_{\rm F}$ and at $-k_{\rm F}$:
$Q_\pm^{\rm (0)}=\int_{-L/2}^{L/2}\rho_\pm^{\rm (0)}(x,t)$,
where
$Q_\pm^{\rm (0)}={1\over 2}(N\pm J)$.
$Q_\pm^{\rm (0)}$ and $Q_\pm$ are related
through the  matrix ${\bf \Omega}$:
$\l[\ba{r}Q_{+}\\ Q_{-}\ea\r]
={\bf \Omega}
\l[\ba{r}Q_{+}^{\rm (0)}\\ Q_{-}^{\rm (0)}\ea\r]$,
where
${\bf \Omega}={1\over 2}\l[\ba{rr}
1+g & 1-g \\
1-g & 1+g
\ea\r]$.

\bibitem{alek1}
A similar calculation for the ground state
can be found in 
A.Yu. Alekseev, V.V. Cheianov and J. Froehlich,
Phys. Rev. {\bf B54} R17320 (1996).

\bibitem{alek2}
A.Yu. Alekseev, V.V. Cheianov and J. Froehlich,
cond-mat/9706061; Phys. Rev. Lett. {\bf 81},
3503 (1998).

\bibitem{safi2}
I. Safi, Eur. Phys. J. {\bf B12}, 451 (1999).

\bibitem{egger2}
R. Egger and H. Grabert, Phys. Rev. Lett. {\bf 79}, 3463 (1997);
R. Egger and H. Grabert, 
Phys. Rev. {\bf B58}, 10761 (1998).

\bibitem{kane}
C.L. Kane and M.P.A. Fisher, Phys. Rev. Lett.
{\bf 68}, 1220 (1992).

\bibitem{fn3}
In this case the crossover from weak
tunneling to ohmic regime would be worth mentioning
\cite{chamon}.
The idea is to map the system to an equivalent 
quasiparticle$\between$electron tunneling duality model
\cite{kane,schmid}
at effective filling factor $\nu_{\rm eff}$.
The system ($=$sample$+$reservoirs) is first
modelized as a chiral electron tunneling model
with different 
fictitious
filling factors, i.e.,
$\nu_1=g$ and $\nu_2=1$. Then a rotation 
in the space of bosonic field is applied.
$\nu_{\rm eff}$ is obtained by comparing the scaling
dimension of electron tunneling operators
in the original and rotated models:
${1\over\nu_1}+{1\over\nu_2}={2\over\nu_{\rm eff}}$.
In the strong electron tunneling (ohmic) limit
we have a boundary condition
$I=\nu_{\rm eff}{e^2\over h}(V_S-V_-)$,
$I=\nu_{\rm eff}{e^2\over h}(V_+-V_D)$.
Together with a voltage drop equation:
$I=\nu{e^2\over h}(V_+-V_-)$
this amounts to the boundary condition (\ref{bc0}).

\bibitem{kawab}
A. Kawabata, J. Phys. Soc. Japan
{\bf 65}, 30 (1996).

\bibitem{finkel}
Y. Oreg and A.M. Finkel'stein,
Phys. Rev. {\bf B54}, R14265 (1996).

\bibitem{glaz}
M.P.A. Fisher and L.I. Glazman, in
{\it Mesoscopic Electron Transport},
edited by L.L. Sohn, L.P. Kouwenhouven and G. Schoen, 
NATO series E, Vol. {\bf 345}, 331 
(Kluwer Academic Publishing, Dordrecht, 1997).

\bibitem{trau} 
B. Trauzettel, R. Egger and H. Grabert,
cond-mat/0109022.

\bibitem{bena} 
C. Bena, S. Vishveshwara, L. Balents, and 
M.P.A. Fisher, cond-mat/0008188.

\bibitem{pono2}
V.V. Ponomarenko and N. Nagaosa, Solid State Commun. 
{\bf 110}, 321, (1999).

\bibitem{chiral}
For $N_R$ ($N_L$) point-like contacts connected to
the right (left) reservoir, the two-terminal conductance 
is given in the strong-coupling for each contact as
\[G_{\rm SD}
=\nu\ {e^2\over h}
{\l[1-\l(-{1-\nu\over 1+\nu}\r)^{N_L}\r]
\l[1-\l(-{1-\nu\over 1+\nu}\r)^{N_R}\r]
\over
1-\l(-{1-\nu\over 1+\nu}\r)^{N_L+N_R}}.
\]

\bibitem{furusaki}
A. Furusaki and N. Nagaosa,
Phys. Rev. {\bf B47}, 4631 (1993).

\bibitem{schmid}
A. Schmid,  Phys. Rev. Lett. {\bf 51}, 1506 (1983).

\end{multicols}
\end{document}